\documentclass[aps,pra,letterpaper,superscriptaddress
,nofootinbib
,twocolumn
,showpacs]{revtex4}

\usepackage{amstext,amsfonts,amsmath,amssymb,ulem,bbm}
\usepackage{color}

\usepackage{txfonts}
\usepackage{algorithm}
\floatname{algorithm}{Protocol}
\usepackage{ftnxtra}
\usepackage[T1]{fontenc}
\usepackage[utf8]{inputenc}
\usepackage{graphicx}
\usepackage[colorlinks]{hyperref}

\newcommand{\beq}{\begin{equation}}
\newcommand{\eeq}{\end{equation}}
\newcommand{\ket} [1] {\vert #1 \rangle}
\newcommand{\bra} [1] {\langle #1 \vert}

\newcommand{\mean}[1]{\langle #1 \rangle}

\DeclareMathOperator\tr{tr}
\DeclareMathOperator\erf{erf}
\newcommand{\ba}{\begin{align}}
\newcommand{\ea}{\end{align}}
\newcommand{\bea}{\begin{eqnarray}}
\newcommand{\eea}{\end{eqnarray}}

\newcommand{\smallfrac}[2]{ \mbox{ $\frac{#1}{#2}$ } }

\newcommand{\bs}[1]{\boldsymbol{#1}}
\newcommand{\GS}{\text{GS}}
\newcommand{\GSvac}{\text{GS}_{\rm vac}}
\newcommand{\GSs}{\text{GS}_{\rm sq}}
\newcommand{\logic}{{\lset}}
\newcommand{\mitigate}{ {\rm cool} }

\newcommand{\primpath}{{\mathcal P}}
\newcommand{\dualpath}{{\tilde{\primpath}}}

\newcommand{\abs}[1]{\left\lvert{#1}\right\rvert}
\newcommand{\abss}[1]{\lvert{#1}\rvert}
\newcommand{\avg}[1]{\left\langle {#1} \right\rangle}
\newcommand{\avgg}[1]{\langle {#1} \rangle}

\newcommand\sqp{r}

\DeclareMathOperator{\cov}{cov}
\DeclareMathOperator{\var}{var}
\DeclareMathOperator{\erfc}{erfc}
\renewcommand{\Re}{\operatorname{Re}}
\renewcommand{\Im}{\operatorname{Im}}

\DeclareMathOperator{\sech}{sech}

\def\natnums{\mathbb{N_+}}

\newcommand{\lset}{{\mathcal L}}

\makeatletter
\@ifundefined{textcolor}{}
{%
 \definecolor{BLACK}{gray}{0}
 \definecolor{WHITE}{gray}{1}
 \definecolor{RED}{rgb}{.9,0,0}
 \definecolor{GREEN}{rgb}{0,.6,0}
 \definecolor{BLUE}{rgb}{0,0,.9}
 \definecolor{CYAN}{cmyk}{1,0,0,0}
 \definecolor{MAGENTA}{cmyk}{0,1,0,0}
 \definecolor{YELLOW}{cmyk}{0,0,1,0}
 }

\makeatother

\usepackage{bm}
\usepackage{mathtools}

\def\reals{\mathbb{R}}
\def\complex{\mathbb{C}}
\def\integers{\mathbb{Z}}

\def\op#1{\hat{#1}}
\def\opvec#1{\op{\vec{#1}}}
\def\opmat#1{\op{\mat{#1}}}
\def\tildevec#1{\tilde{\vec{#1}}}

\def\id{I}

\def\1{\mat{\id}}

\def\mat#1{\mathbf{#1}}

\renewcommand{\vec}[1]{\bm{\mathrm{#1}}}

\def\tp{\mathrm{T}}
\def\herm{\mathrm{H}}

\renewcommand{\sout}[1]{}

\newcommand{\blk}{\protect\color{BLACK}}

\allowdisplaybreaks

\graphicspath{{graphics/}}

\begin{document} 

\title{Anonymous broadcasting of classical information with a continuous-variable topological quantum code }

\author{Nicolas C. Menicucci}
\affiliation{Centre for Quantum Computation and Communication Technology, School of Science, RMIT University, Melbourne, Victoria 3001, Australia}
\affiliation{School of Physics, The University of Sydney, Sydney, NSW 2006, Australia}

\author{Ben Q. Baragiola}
\affiliation{Centre for Quantum Computation and Communication Technology, School of Science, RMIT University, Melbourne, Victoria 3001, Australia}
\affiliation{Centre for Engineered Quantum Systems, Department of Physics and Astronomy, Macquarie University, North Ryde, NSW 2109, Australia}

\author{Tommaso F. Demarie} 
\affiliation{Singapore University of Technology and Design, 8 Somapah Road, Singapore 487372}
\affiliation{Centre for Engineered Quantum Systems, Department of Physics and Astronomy, Macquarie University, North Ryde, NSW 2109, Australia}
\affiliation{Centre for Quantum Technologies, National University of Singapore, Block S15, 3 Science Drive 2, Singapore 117542}

\author{Gavin K. Brennen}
\affiliation{Centre for Engineered Quantum Systems, Department of Physics and Astronomy, Macquarie University, North Ryde, NSW 2109, Australia}
\begin{abstract}
Broadcasting information anonymously becomes more difficult as surveillance technology improves, but remarkably, quantum protocols exist that enable provably traceless broadcasting. The difficulty is making scalable entangled resource states that are robust to errors. We propose an anonymous broadcasting protocol that uses a continuous-variable surface-code state that can be produced using current technology. High squeezing enables large transmission bandwidth and strong anonymity,  and the topological nature of the state enables local error mitigation.  %
\end{abstract}
\pacs{%
03.67.Dd%
, 03.67.Pp%
, 42.50.Ex%
}

\date{\today}
\maketitle

\section{Introduction}
Almost every aspect of modern society relies on information processing. As digital surveillance capabilities continue to expand, so does demand for guaranteed-anonymous communication strategies. %
An important primitive for privacy-preserving routines is anonymous broadcasting ~\cite{Movahedi:2014tt}, which can facilitate, for example, tipping off the police anonymously, secret balloting, and secure electronic auctions~\cite{Stajano:2000kh}, and anonymous cryptocurrency transactions \cite{T.-Ruffing:2016aa}.
In the original classical formulation~\cite{Chaum:1988dg} and its improvements ~\cite{BT07,BJT10}, $n$ players establish shared keys enabling one party to reveal a single bit of information while keeping her identity secret. The first \emph{quantum} protocol allowing one to communicate classical information anonymously was proposed in Ref. \cite{Boykin:2002PhD}\blk. A more efficient and secure quantum protocol for anonymous quantum and classical broadcasting was reported by Christandl and Wehner in Ref. \cite{Christandl:2005cc}. Here, a trusted resource distributes ahead of time an $n$-partite entangled state
\begin{align}
\label{eq:GHZdef}
	\ket{\text{GHZ}}=\frac{1}{\sqrt{2}} \Bigl(\ket{0_1 \dotsm 0_n}+\ket{1_1 \dotsm 1_n} \Bigr),
\end{align}
one qubit to each party. 
The key feature of this quantum protocol is that it is completely \emph{traceless}---i.e.,~the sender's identity cannot be determined (better than guessing) even if all resources are made public at the end of the protocol. Remarkably, tracelessness cannot be achieved classically.
This protocol and its later improvements~\cite{BBFGT07,Cai:2013}, however, suffer from decoherence from unwanted interactions with the environment. Indeed, the issue of decoherence is rather challenging to overcome, and it has surprisingly been ignored in all previous works.
A solution to this problem is to encode the shared resource in a quantum error-correcting code~\cite{Nielsen2000}. A practical code should be fast to prepare %
and easy to correct using mostly local operations by the players involved. Surface codes~\cite{Kitaev2003} satisfy these requirements. These
have been extensively studied for the purpose of providing sustained quantum memories or for fault-tolerant quantum computation~\cite{Pachos:2012ug}, and recent experiments~\cite{Barends:2014fk} have built small prototype qubit toric codes. However, the overhead in gates and qubits for such quantum processing is daunting~\cite{QURE}.  

Here we show that much simpler tasks for communicating \emph{classical information} benefit from the topological protection of such codes. In particular, we present a protocol for quantum-assisted anonymous broadcasting using a recently developed continuous-variable (CV) toric code~\cite{Demarie:2014jx}. The motivation for using this resource is threefold: (1)~the topological nature of the state allows for error mitigation; (2)~the state can be easily prepared and distributed to the players using Gaussian resources and operations; and (3)~using a CV resource allows for a larger communication bandwidth than either the classical or the discrete quantum counterpart. This bandwidth is limited only by the initial squeezing level in the resource.

\section{Anonymous broadcasting with the qubit toric code}

\begin{figure}
\includegraphics[width=\columnwidth]{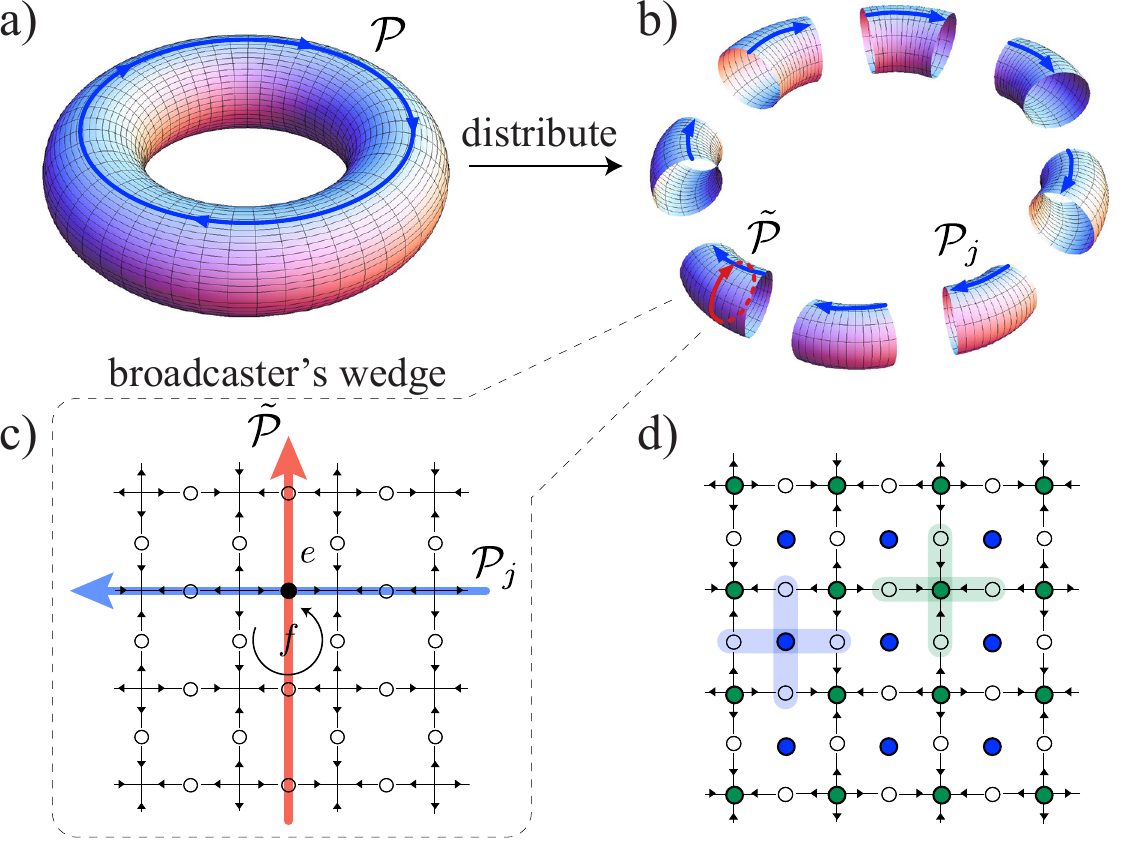}
\caption{Sketch of the protocol.  
(a)~A CV surface-code ground state (with squeezed logical modes) is prepared on a torus. The players decide beforehand to perform measurements along a loop around the torus (shown in blue). (b)~The state is distributed to the players, one wedge to each. (c) Close-up of the lattice on the broadcaster's wedge. Physical bosonic modes are assigned to each edge, and each edge is assigned an orientation. Similarly, the faces are given a uniform orientation (one face is shown for reference).  For the indicated edge $e$, $f(e,\primpath)=+1$ with respect to path $\primpath$ and $f(e, \dualpath)=+1$ with respect to path $\dualpath$ (see Sec.~\ref{subsec:finitelysqueezed}).  The broadcaster performs the unitary displacement, Eq.~(\ref{eq:unitarydisplacement}), on a loop $\dualpath$ around her wedge (shown in red) , which encodes a message $r\in \reals$.  Next, each party~$j$ measures an operator, Eq.~(\ref{Eq:MomWedge}), along an arc $\primpath_j$ of the loop $\primpath$ (shown in blue). The players publicly announce their measurement outcomes $\{m_j\}_{j=1}^n$, and the broadcast message is computed as their (noisy) weighted sum.  (d)~Error mitigation strategy.  Additional blue (green) ancillae are quasi-locally coupled to modes surrounding faces (vertices), which perform error mitigation by dissipative cooling  (see Sec.~\ref{sec:error_mit}).
}
\label{fig:TCandprotocol}
\end{figure}

We illustrate the main idea with a qubit toric code. Consider an $n \times m$ square lattice with a sets of vertices $\mathcal{V}=\{v\}$, faces $\mathcal{F}=\{f\}$, and edges $\mathcal{E}=\{e\}$. The lattice lies on a torus, and there is one qubit logically assigned to each edge. The code states are $+1$ eigenstates of the stabilizers~\cite{Kitaev2003, JP2012}
\begin{align}
	\op A_v &\coloneqq \prod_{e \in +_v}\op X_e
	\qquad \forall v \in \mathcal V
	,
\\
	\op B_f&\coloneqq  \prod_{e \in \square_f}\op Z_e
	\qquad \forall f \in \mathcal F
	.
\end{align}
On the torus, these operators stabilize a 4-dimensional subspace,
which encodes two logical qubits~\cite{Hamma:2005et}. For one of these qubits, the logical $\op Z$ and $\op X$ Pauli operators
are, respectively, the string operators
\begin{align}
	\op Z_\primpath &= \prod_{e\in\primpath}\op Z_e
&\text{and}&
&
	\op X_\dualpath
	&= \prod_{e\in\dualpath}\op X_e
	,
\end{align}
where $\primpath$ is any closed loop along the primal lattice encircling the hole of the torus, and $\dualpath$ is any closed loop along the dual lattice threading through the hole (see Fig.~\ref{fig:TCandprotocol}).\footnote{There is a second set of similarly defined string operators that serve as logical Pauli operators for the second logical qubit~\cite{Kitaev2003}. We only need one logical qubit for the protocol, so we omit their specification to simplify the notation.}

Here and in the following, we always consider a scenario with $n$ participants, of whom exactly one of them, Alice, wants to anonymously broadcast a public message. The broadcast resource is a toric-code state. In particular, we want a multi-qubit state that is simultaneously stabilized by $\op A_v$ $\forall v \in \mathcal V$ and also stabilized by~$\op Z_\primpath$. One choice for this is
\begin{align}
	\ket {\GS_{00}} &\coloneqq \prod_v \frac {1} {\sqrt{2}} (\hat \id+\hat A_v)\ket{0 \dotsm 0}.
\end{align}
The notation `GS' stands for a \emph{ground state} of the toric-code Hamiltonian~\cite{Kitaev2003}, and this means that $\op A_v$ and $\op B_f$ stabilize the state $\forall v \in \mathcal V$ and $\forall f \in \mathcal F$. The subscript~`00' indicates that both logical qubits are prepared in the logical~$\ket 0$ state. The qubits are logically grouped into $n$~wedges, and the wedges are distributed, one to each player (see Fig.~\ref{fig:TCandprotocol}).

When Alice wants to anonymously broadcast the message $r=1$, she performs the string operation $\op X_\dualpath$ around the loop on her wedge [see Fig.~\ref{fig:TCandprotocol}(c)], while for the message $r=0$, she does nothing. Next, each party $j$ measures qubits in the local~$\op Z$ basis along an arc of the wedge and publicly announces the parity $m_j \in \{0,1\}$ of the outcomes. The broadcast message is recovered from the sum ${\sum_{j=1}^n m_j=r \pmod 2}$.

When using a graph with $\abs \dualpath = 1$ (i.e.,~just one qubit wide, a loop along $\primpath$), then $\ket{\GS_{00}}$ is just a GHZ state in the $\ket \pm$ basis. For such a torus (loop), %
vertex stabilizers reduce to pairs of adjacent Paulis ($\hat X \otimes \hat X$) along $\dualpath$, and face stabilizers do not exist. In either case (GHZ or full toric code), the variance of any individual party's measurement is maximal, and no collusion by any proper subset of the non-broadcasting players will reveal any information about the identity of the broadcaster. 

Using a qudit toric code~\cite{Bullock:2007ih} (or qudit GHZ state) the protocol generalizes to allow a single party to anonymously broadcast any $d$-ary integer~$r \in \integers_d$ %
by applying the string operator~$\op X^r_\dualpath$, where $\op X$ represents the Weyl-Heisenberg shift operator$\pmod d$. Then, $\sum_{j=1}^n m_j = r \pmod d$. This amounts to broadcasting $\log_2 d$ bits of data per round. Alternatively, such a protocol could instead allow up to $d-1$ broadcasters (out of the $n$ total players) to signal `yes' %
by each applying~$X_\dualpath$ around their own wedge. In that case, $\sum_{j=1}^n m_j$ would return the number of `yes' broadcasters$\pmod d$.

The advantage of using a toric-code state instead of a simple GHZ state appears when one considers noise (errors) in the protocol. Notably, since errors in the surface code can be diagnosed by measuring stabilizers, almost all such measurements and corrections are local to each party and can be corrected without disrupting the protocol~\cite{Fowler:2012gc,Anwar:2014vr}. The exceptions are those stabilizers that straddle the boundary between wedges, and these may be measured with the assistance of Bell pairs shared between nearest-neighbor players to enable nonlocal gates \cite{Brennen:2003ba}. To do this, the number of entangled pairs needed grows as the number of players and as the width of each wedge. This width, as shown in Appendix~\ref{sec:width}, is a small constant.

\section{Anonymous broadcasting with a continuous-variable toric code}

\subsection{Finitely squeezed continuous-variable surface codes}	
\label{subsec:finitelysqueezed}

The ideal CV surface code~\cite{Zhang2008b} is a straightforward generalization of the qudit surface code, but it represents an unphysical model because the required states are infinitely squeezed. A finitely squeezed CV surface code is an experimentally accessible, physical approximation of this code~\cite{Demarie:2014jx}. This model starts with an $n \times m$ square lattice with a sets of vertices $\mathcal{V}=\{v\}$, oriented faces $\mathcal{F}=\{f\}$, and oriented edges $\mathcal{E}=\{e\}$ just like in the qudit case~\cite{Bullock:2007ih,Demarie:2014jx}. An independent, local bosonic mode is logically assigned to each edge of the lattice, for a total of $2  n  m$ modes, with quadrature operators $\op q_e, \op p_e$ obeying $[\op q_e,\op p_{e'}] = i\delta_{e,e'}$ (${\hbar=1}$). %

A finitely squeezed CV surface code is the nullspace of a set of nullifiers, $\{ \op{\eta}_i \}$, defined on each vertex and face of the lattice (i.e., $\forall i \in \mathcal V \cup \mathcal F$).  For a given local, mode-wise squeezing factor~$s$, a finitely squeezed CV surface code is not unique; see Appendix~\ref{sec:covmat}. We choose to describe relevant features of a CV surface code using the symmetric nullifiers because they are conceptually simpler.  These nullifiers are
	\begin{subequations} \label{eq:nullifiers}
	\begin{align}
	\hat{\eta}_v &\coloneqq \frac{1}{\sqrt{8}}\sum_{e \in +_v} \left(s\op q_{e}+is^{-1} \op p_{e}\right)
	& \forall v &\in \mathcal V,  \\
	\hat{\eta}_f &\coloneqq \frac{1}{\sqrt{8}}\sum_{e \in\square_f} o(e,f) \left(s\op p_{e}-is^{-1}\op q_{e} \right)
	& \forall f &\in \mathcal F, 
	\end{align}
	\end{subequations}
where the orientation sign factor ${o(e,f) = \pm 1}$ if edge $e$ is oriented the same (opposite) as face $f$. The nullifiers satisfy the commutation relations
\begin{align}
	[\hat{\eta}_v,\hat{\eta}_{v'}] = [\hat{\eta}_f,\hat{\eta}_{f'}] = [\hat{\eta}_v,\hat{\eta}_{f}] = [\hat{\eta}_v,\hat{\eta}^{\dagger}_{f}] = 0
\end{align}
$\forall v \in \mathcal V$ and $\forall f \in \mathcal F$. As a consequence of finite squeezing, the nullifiers are not Hermitian (whereas they are so in the infinitely squeezed case~\cite{Zhang2008b}). This makes them fail to commute with their conjugates when the two share an edge:
\begin{subequations}
\label{eq:nullnoncommute}
\begin{align}
	[\hat{\eta}_v,\hat{\eta}^{\dagger}_{v'}] &\neq0 \qquad \forall (v,v') \in \mathcal E
	,
\\
	[\hat{\eta}_f,\hat{\eta}^{\dagger}_{f'}]&\neq 0 \qquad \forall (f \cap f') \in \mathcal E.
\end{align}
\end{subequations}

By definition, a CV surface-code state~$\ket{\GS}$ is any state that satisfies
\begin{align}
	\op \eta_v \ket \GS
&=
	\op \eta_f \ket \GS =0 \qquad \forall v\in\mathcal{V}, \forall f\in\mathcal{F}. 
\end{align}
Note that we have again used the notation~`GS' to indicate that any such state is a \emph{ground state} of a  
CV surface-code Hamiltonian~\cite{Demarie:2014jx}.

It will turn out that the (related, but inequivalent) CV surface-code state that results from measuring a CV cluster state~\cite{Demarie:2014jx} will be easier to work with for our explicit calculations. The differences between this and a symmetric CV surface-code state, along with all explicit details of the construction of the required states and their measurement statistics used for this work, is given in Appendix~\ref{sec:covmat}. 
On the torus there are only ${nm-1}$ independent vertex nullifiers and ${nm-1}$ independent face nullifiers. Hence, the nullifiers do not span the space of physical modes. And analogous to the two logical qubits encoded in the qubit toric code~\cite{Pachos:2012ug}, there are two unconstrained, topological, harmonic-oscillator modes in the CV toric code. These two \emph{logical modes}, which define a two-mode Hilbert space~$\mathcal H_\logic$, are entirely nonlocal and are independent of the squeezing. Since the nullifiers span a $(2nm-2)$-mode Hilbert space~$\mathcal H_{\rm null}$, the logical modes and the nullifiers together span the full Hilbert space of the $2nm$ local modes.

The projector onto the toric-code logical subspace is
\begin{align}
	\opmat P_\logic \coloneqq \ket{\vec{\eta}} \bra{\vec{\eta}}_{\rm null} \otimes \op{\mat \id}_\logic,
\end{align}
where the tensor-product decomposition is $\mathcal H_{\rm null} \otimes \mathcal H_\logic$, and where $\ket{\vec{\eta}}$ is the simultaneous zero eigenstate of all the nullifiers%
.
We define the two-mode \emph{logical vacuum state}~${(\ket 0 \otimes \ket 0)_\logic}$ as the restriction of the vacuum state of all local modes to the two-mode logical Hilbert space:
\begin{align}
\label{eq:logicalvac}
	\Bigl( \ket 0 \bra 0 \otimes \ket 0 \bra 0 \Bigr)_\logic \coloneqq \tr_{\rm null} \left[ \ket 0 \bra 0^{\otimes(2nm)} \right].
\end{align}
This state is pure because we define the mode transformation from local modes to $\mathcal H_{\rm null} \otimes \mathcal H_\logic$ to be passive (total number conserving).
A full description of a finitely squeezed CV toric code goes beyond the scope of this work and will be presented elsewhere. 

In the mean time, there are two important states we must identify for our work. We include these below, with their derivation given in Appendix~\ref{sec:logicalmodes}.
The first is the 
\emph{toric-code logical vacuum state}
	\begin{align}
	\label{eq:GSvac}
		\ket{\GSvac} \coloneqq  \ket{\vec \eta}_{\rm null} \otimes \Bigl( \ket 0 \otimes \ket 0 \Bigr)_\logic \;,
	\end{align}
which will be used to demonstrate a proof-of-principle error mitigation strategy in Sec.~\ref{sec:error_mit}.  The second is the state that results from preparing a CV toric-code state by measuring a CV cluster state~\cite{Demarie:2014jx}. This state, which we call the \emph{toric-code logical squeezed state}, will be used to analyze the anonymous broadcasting protocol below. It has the form
	\begin{align}
	\label{eq:GSs}
		\ket{\GSs} \coloneqq  \ket{\tildevec \eta}_{\rm null} \otimes \Bigl( \ket{0;s} \otimes \ket{0;s} \Bigr)_\logic \; ,
	\end{align}
where $\ket{0;s}$ is a momentum-squeezed vacuum state with squeezing factor~$s$. Nevertheless, it is still a ground state (hence, `GS') of a CV toric-code Hamiltonian~\cite{Demarie:2014jx}. The nullifiers used to define~$\ket\GSs$ are slightly different from the symmetric nullifiers shown in Eqs.~\eqref{eq:nullifiers}---hence the tilde on~$\tildevec \eta$%
. The logical subspace, however, is exactly the same in both cases. (For further information, see Appendix~\ref{sec:covmat}.)

The reasons we use this state, despite the aforementioned complications, are (1)~we know how to make it from a large-scale CV cluster state~\cite{Demarie:2014jx}, (2)~large-scale CV cluster states have been demonstrated experimentally (see Sec.~\ref{sec:implementation}), and (3)~the covariance matrix for this state has a $pp$ submatrix that is of a particularly simple form, which simplifies the analysis of its performance for anonymous broadcasting (see Appendix~\ref{sec:covmat}).

For completeness, we note that in standard quantum-optics language~\cite{Walls2008},
\begin{align}
	\ket{0;s}
&
\coloneqq
	\op S(-\ln s) \ket 0
	,
&
	\op S(\xi)
&
\coloneqq
	\exp \left[\smallfrac 1 2( \xi^* \op a^2 - \xi \op a^{\dag 2} ) \right]
	,
\end{align}
where ${\xi = -\ln s}$~is the squeezing parameter. Thus, with our conventions, we have for any single mode
\begin{align}
	\bra{0;s} \op q^2 \ket{0;s} &= \frac {s^2} {2}
	,
&
	\bra{0;s} \op p^2 \ket{0;s} &= \frac {1} {2s^2}
	.
\end{align}
The case $s=1$ corresponds to the ordinary vacuum state.

\begin{algorithm}[t]
\caption{Finite-squeezing CV anonymous broadcasting}
 \label{prot:CVAB}
~\\
{\bf Steps of the protocol:}
\begin{enumerate}
\item {\bf Initialization:}
 A CV toric-code logical squeezed state~$\ket{\GSs}$ is prepared [Eq.~\eqref{eq:GSs}].  The state is distributed, one wedge to each player. 
\item {\bf Broadcasting:}\\
To anonymously broadcast the real number $r$, Alice performs the displacement $\op D_r$ [Eq.~(\ref{eq:unitarydisplacement})] on her wedge.
\item {\bf Local measurements:}\\
	Each player measures her portion of the string momentum, $\op M_j$ [Eq.~(\ref{Eq:MomWedge})], and records the outcome $m_j\in \reals$.
\item {\bf Determining the broadcast message:} \\
All players publicly announce their results $\{m_j\}$. The message broadcast by Alice can be inferred from the noisy weighted sum $M$ in Eq.~(\ref{eq:inferredmessage}).
\end{enumerate}
\end{algorithm}

	\subsection{Anonymous broadcasting protocol}
	\label{subsec:abprotocol}
	
Given a CV toric code, the anonymous-broadcasting protocol is summarized in Protocol~\ref{prot:CVAB} and graphically represented in Fig.~\ref{fig:TCandprotocol}.  We make use of a non-local string momentum operator
	\begin{align} \label{eq:stringmomentum}
		\op M\coloneqq \frac {1} {\sqrt{\abs{\primpath}}} \sum_{e\in \primpath} f(e, \primpath)\op p_e. %
	 \end{align}
where $\primpath$ is a loop around on the primal lattice. For each edge the orientation factor $f(e, \primpath) = \pm 1$ if the edge has the same (opposite) orientation as the path $\primpath$. 
For the toric-code logical squeezed state $\ket{\GSs}$, the variance of the string momentum operator~$\hat M$ is $(\Delta M)^2=\frac{1}{2s^2}$, with $\langle \hat M \rangle = 0$, as shown in Appendix~\ref{sec:covmat}.  The torus is divided into $n$ wedges, and each is distributed to a single player.
To broadcast the real number $r$, Alice wishes to perform a displacement of the string momentum $\op M \mapsto \hat M + r$ by means that are not detectable once the measurements have begun~\cite{Christandl:2005cc}. To this end, she applies a displacement on the dual lattice along the loop $\dualpath$ by applying the unitary
\begin{align} \label{eq:unitarydisplacement}
\op D_r = \exp \left( i r \sqrt{ | \dualpath | }  \sum_{e\in \dualpath} f(e, \dualpath) \op q_e \right)
\end{align}
on her wedge. Here,  $f(e, \dualpath)=\pm 1$ if the edge $e$ has the same (opposite) direction as the framing of the path $\dualpath$, where the framing of a path is to the right and normal to its direction [see Fig.~\ref{fig:TCandprotocol}(b)]. 

After the broadcasting stage of the protocol, the string momentum operator $\op M$ is measured, with each player contributing a measurement on her wedge. The party holding wedge $j \in \{1,2,\dots,n\}$ measures her portion of the string momentum operator,
\begin{align} \label{Eq:MomWedge}
	\op M_j \coloneqq \frac{1}{\sqrt{\abss{\primpath_j}}} \sum_{e\in \primpath_j}o(e)\op p_e,
\end{align}
along an arc $\primpath_j$ of the loop $\primpath$. Each party records the outcome $m_j \in \reals$. 
During the measurements, the path $\primpath = \bigcup_{j=1}^n \primpath_j$ must be a closed loop. This implies pre-agreement between the players and active classical communication during the protocol to establish a different connected path in case of errors at the wedge boundaries. 

In the final step of the protocol, all players publicly announce their measurement results $\{m_j\}$. The broadcast message is recovered by calculating the noisy, weighted sum,
\begin{align} \label{eq:inferredmessage}
	M = \frac{1}{ \sqrt{ \abs{\primpath} } } \sum_{j=1}^n \sqrt{\abss{\primpath_j} } \, m_j,
\end{align}
which is a classical random variable with mean~$r$ and variance~$(\Delta M)^2 = \frac {1} {2s^2}$, as shown in Appendix~\ref{sec:covmat}. \footnote{We have assumed, without loss of generality, that the face and edge orientation at the edge $e_A$ of the intersection of the arc $\primpath({\rm Alice})$ and the loop $\dualpath$ satisfies $(-1)^{f(e_A)+o(e_A)}=1$; otherwise, $r$ acquires that sign.}

		\section{Broadcast channel capacity} \label{sec:capacity}

In this section, we calculate the channel capacity for the broadcast protocol discussed above. %
Since the message space is unbounded, the capacity is technically infinite. Therefore, in order to get a finite quantity, we will calculate the channel capacity conditioned on a fixed variance~$\tau^2$ of the message to be broadcast. (This does not specify the shape of the broadcast message distribution, of course, since two possibilities would be a Gaussian with variance~$\tau^2$ and a binary distribution with $\delta$-function support only at $\pm \tau$.) The result presented here was first calculated by Shannon~\cite{Shannon:1949fb,Shannon:1959ge}. We include our own derivation in order to maintain a self-contained presentation and because it is straightforward and rather elegant.

For an input broadcast message ${R\in \reals}$ and some output reconstructed message ${M \in \reals}$, the variance-restricted channel capacity is  $C=\max_{p_R(r)}I(R;M)$, where the maximum is over all input probability distributions $p_R(r)$ with variance~$\tau^2$, and $I(R;M)=H(M)-H(M|R)$ is the mutual information between $R$ and $M$~\cite{Cover:2012ub}. The conditional probability $p_{M|R}(m|r) = N_{m,(\Delta M)^2}(r)$ is a normal distribution  [see Eq.~\eqref{eq:Gauss}]  in output $m$ with mean $r$ and variance $(\Delta M)^2$ from Eq.~\eqref{eq:DeltaM}.

For an arbitrarily distributed~$R$ with mean~$\mu$ and variance~$\tau^2$, the cumulant vector~\cite{Jaynes:2003vj} for $R$ is ${\vec c_R = (\mu, \tau^2, c_{3}, c_{4}, \dotsc)}$, and that for $M$ is called $\vec c_M$. %
Using the law of total probability,
\begin{align}
	p_M(m) &= \int dr\, p_{M|R}(m|r) p_R(r) \nonumber \\
	&= (N_{0,(\Delta M)^2} * p_R)(m)\,,
\end{align}
where $*$ indicates convolution. Cumulants add under convolution~\cite{Jaynes:2003vj}. Therefore,
\begin{align}
	\vec c_M &= \vec c_R + \bigl(0, (\Delta M)^2, 0, \dotsc \bigr) \nonumber \\
	&= \bigl(\mu, {\tau^2+(\Delta M)^2}, c_{3}, c_{4}, \dotsc \bigr)\,.
\end{align}

Note that $H(M|R)$ is fixed by the channel since $p_{M|R}(m|r)$
is a function only of $(m-r)$, and thus averaging over $R$ does not change the entropy. Therefore, the only difference that $p_R$ makes to $I(R;M)$ is through $H(M)$. We can maximize $I(R;M)$ by maximizing $H(M)$ (subject to the $\tau^2$ constraint), which means requiring that $p_M$ be Gaussian (see Appendix~\ref{sec:math}) with variance $\tau^2 + (\Delta M)^2$ and arbitrary mean. This can be achieved by requiring all cumulants beyond the second of~$\vec c_M$ to be zero---i.e., $\vec c_M = (\mu, \tau^2+(\Delta M)^2, 0, 0, \dotsc)$. Therefore, $\vec c_R = (\mu, \tau^2, 0, 0, \dotsc)$, which means that the maximizing $p_R$ is also Gaussian. For a given variance~$\tau^2$ of the message, this choice maximizes the mutual information and thus defines the (variance-restricted) channel capacity (see Appendix~\ref{sec:math}):%
\begin{align} \
	C &= \frac 1 2 \log\bigl[2\pi e \bigl(\tau^2+(\Delta M)^2\bigr)\bigr] - \frac 1 2 \log \bigl[2\pi e (\Delta M)^2 \bigr] \nonumber \\
	&= \frac 1 2 \log \left(1 + \alpha \right)\,, \label{eq:channelcapacity}
\end{align}
where the signal-to-noise ratio (SNR) of the broadcast is
	\begin{align} \label{eq:SNR}
		\alpha = \frac {\tau^2} {(\Delta M)^2}.
	\end{align}
There exist lattice codes for sending digital information through such a channel that achieve this capacity~\cite{Urbanke:1998fm}.

	\section{Broadcaster anonymity} \label{sec:anonymity}

Due to finite squeezing the broadcast will not be completely anonymous. We precisely quantify the tradeoff between anonymity and channel capacity in terms of squeezing, and hence signal-to-noise ratio~(SNR). We first discuss anonymity: this is predicated on the assumed inability to identify the broadcaster based on the local measurement outcomes. The degree to which this is true depends on the SNR of the message strength to the noise in the local measurement. A high degree of anonymity depends on this being small. However, the signal strength cannot be too small lest the broadcast be too weak to be detected. %

In this section we quantify the anonymity of the broadcast channel in terms of how much information about the identity of the sender leaks out into the classical measurement record. 
First, we need the measurement covariance matrix shared among the players prior to the broadcast.  This is done for various cases in Appendix~\ref{sec:covmat}, including the CV toric code as well as simpler graphs such as the CV GHZ state and the open boundary CV surface code.
We assume a surface-code state with toroidal boundary conditions, as discussed in Appendix~\ref{subsec:toric}, in order to simplify the calculation by putting all players on the same footing. A similar calculation is possible using other boundary conditions and more general assumptions, but our purpose is simply to quantify the amount of anonymity in a basic instance of the protocol.%

		\subsection{Players' covariance matrix after broadcast}

In Appendix~\ref{subsec:toric}, we calculate the covariance matrix of the players' individual measurement outcomes before any broadcast is made,  given by Eqs.~(\ref{eq:playervariance}--\ref{eq:playercovariance}). %
The full covariance matrix for the random measurement-results vector~${\vec M}$ can be written using the definition for the circulant matrix in Appendix~\ref{sec:math}, Eq.~\eqref{eq:Cn}%
:
\begin{align}
\label{eq:meascov}
	\mat \Sigma \coloneqq \avg{ {\vec M}  {\vec M}^\tp} = \frac {-s^2} {2w} \mat C_n\left(- \frac {w} {s^4} - 2 \right)\,,
\end{align}
 where $w$~is the width of each wedge.

Let the identity of Alice (the broadcaster) be associated with a random variable $A \in \{1, \dotsc, n\}$. (It is random because other people wishing to discover her identity do not know who she is.) We assume that she wishes to broadcast a real number~$r \in \reals$, which we shall treat as an instantiation of a Gaussian-distributed random variable~$R \sim N_{0,{\tau^2}}(r)$, as is prescribed to be optimal in Sec.~\ref{sec:capacity}. Conditioned on Alice actually being player~$a$ and applying the string-momentum shift along~$\dualpath$  to implement the broadcast, the actual random measurement outcome for each player can be written 
\begin{align}
\label{eq:shift}
	M_{j|a} \coloneqq %
	 M_j + \sqrt n R \delta_{ja}\,,
\end{align}
since $n = \abss{\primpath}/\abss{\primpath_j}$.
Then, the variance and covariance of the actual measurement outcomes \emph{when averaged over the actual message sent} are, respectively,
\begin{align}
	\avgg{M_{j|a}^2} &= \frac {1} {2s^2} + \frac {s^2} {w} + n {\tau^2} \delta_{ja}\,, \\
	\avgg{M_{j|a} M_{j\pm1|a}} &= \frac {-s^2} {2w}\,.
\end{align}
This gives the following covariance matrix of the actual random vector of outcomes, conditioned on the broadcaster being player~$a$:
\begin{align}
	\mat \Sigma_{|a} &\coloneqq \avg{\vec M_{|a} \vec M_{|a}^\tp} = {\mat \Sigma} + n {\tau^2} \mat e_{aa}\,,
\end{align}
where $\mat e_{aa}$ is a matrix with a~1 in the $(a,a)$ entry and zeros everywhere else.
		\subsection{Information leakage about broadcaster's identity}

We model the leakage of information about the broadcaster's identity in terms of the mutual information~$I(\vec M;A)$ between the random vector of measurement outcomes~$\vec M$ (averaged over the broadcaster~$A$ and the message~$R$) and the random variable~$A$ identifying the broadcaster~\cite{Cover:2012ub}. In other words, how much information about~$A$ can be extracted from~$\vec M$? More specifically, this measures how much the entropy of~$A$ is reduced (on average) if one has access to the measurement record~$\vec M$:
\begin{align}
	I(\vec M;A) = H(A) - H(A|\vec M)\,.
\end{align}
Symmetry of the mutual information means that we can also write it as
\begin{align}
	I(\vec M;A) = H(\vec M) - H(\vec M|A)\,,
\end{align}
which will be more straightforward to calculate.

The conditional entropy is the entropy of~$\vec M$ if one knows who the broadcaster is, averaged over both the message and the broadcaster's identity:
\begin{align}
	H(\vec M|A) = \avg{-\log p_{\vec M|A}(\vec M|A)}_{\vec M,A}\,.
\end{align}
We assume, for simplicity, that we have no initial information about the broadcaster's identity---a flat prior over all possible broadcasters:
\begin{align}
	A \sim p_A(a) = \frac 1 n\,.
\end{align}
From the subsection above, we know the distribution of the message~$\vec M_{|a}$ conditioned on knowing who the broadcaster is:
\begin{align}
	\vec M_{|a} \sim p_{\vec M|A}(\vec m|a) = N_{\vec 0, \mat \Sigma_{|a}}(\vec m),
\end{align}
where we used the notation for a multivariate Gaussian from Eq.~\eqref{eq:Gaussmulti}.  Therefore (see Appendix~\ref{sec:math}),
\begin{align}
\label{eq:HMgivenA}
	H(\vec M|A) &= \avg{\frac 1 2 \log \det \left(2\pi e \mat \Sigma_{|A} \right) }_A \nonumber \\
	&= \frac 1 2 \log_2 \det \left[ 2\pi e \left( \mat \Sigma + n {\tau^2} \mat e_{1,1} \right) \right]\,.
\end{align}
Note that $n {\tau^2}$ could have just as well been added to any other location on the diagonal; the (1,1) entry was chosen by fiat.

Using the law of total probability, we can calculate
\begin{align}
	\vec M \sim p_{\vec M}(\vec m) &= \sum_{a=1}^n p_{\vec M|A}(\vec m|a) p_A(a) \nonumber \\
	&
	= \frac 1 n \sum_{a=1}^n N_{\vec 0, \mat \Sigma_{|a}}(\vec m)\,.
\end{align}
This is not a Gaussian; rather, it is a mixture of Gaussians with different covariance matrices. Nevertheless, we can use the law of total expectation to calculate the post-measurement covariance matrix:
\begin{align}
%
	\avgg{\vec M \vec M^\tp}_{\vec M} &= \frac 1 n \sum_{a=1}^n \avgg {\vec M_{|a} \vec M_{|a}^\tp}_{\vec M|a} \nonumber \\
	&= \frac 1 n \sum_{a=1}^n \mat \Sigma_{|a} \nonumber \\
	&= \mat \Sigma + {\tau^2} \mat \id
\end{align}
By Eq.~\eqref{eq:entupper} in Appendix~\ref{sec:math}, we can use this to place an upper bound on $H(\vec M)$:
\begin{align}
\label{eq:HMbound}
	H(\vec M) \leq \frac 1 2 \log \det \left[2\pi e \left( \mat \Sigma + {\tau^2} \mat \id \right) \right]\,.
\end{align}
And hence, combining Eqs.~\eqref{eq:HMgivenA} and~\eqref{eq:HMbound}, we have
\begin{align} \label{eq:mutual}
	I(\vec M;A) \leq \frac 1 2 \log \left[ \frac
	{\det \bigl(\mat \Sigma + {\tau^2} \mat \id \bigr)}
	{\det \bigl(\mat \Sigma+ n {\tau^2} \mat e_{1,1} \bigr)}
	\right]\,.
\end{align}
For convenience we define
\begin{align}
\label{eq:epsilon}
	\epsilon = \frac {(\Delta M)^2} {(\Delta M_j)^2 - (\Delta M)^2}, 
\end{align}
such that the quantities that appear in Eq.~(\ref{eq:mutual}) can be written
\begin{align}
	\mat \Sigma + {\tau^2} \mat \id &= \frac {-s^2} {2w} \mat C_n[-2(1+\epsilon+\epsilon \alpha)]\,, \\
	\mat \Sigma + n {\tau^2} \mat e_{1,1} &= \frac {-s^2} {2w} \mat C_n[-2(1+\epsilon),-2n \epsilon \alpha]\,,
\end{align}
where $\alpha$ is the SNR given in Eq.~(\ref{eq:SNR}).
Using Eqs.~\eqref{eq:detCn}, and~\eqref{eq:detCshift}, we obtain an explicit bound on the amount of information about the broadcaster's identity leaked within the measurement outcomes (assuming $n \ge 3$):
\begin{align}
\label{eq:mutinfobound}
	I(\vec M;A) \leq \frac 1 2 \log \left\{ \frac
	{T_n \left(1+\epsilon+\epsilon \alpha \right) -1}
	{\left(1+\epsilon \alpha \frac {\partial} {\partial \epsilon} \right) \left[T_n \left(1+\epsilon \right) -1 \right]}
	\right\}\,.
\end{align}
where $T_n$ is the $n$th-order Chebyshev polynomial of the first kind, valid for $n\ge 3$. 
The mathematical form of Eq.~\eqref{eq:mutinfobound} can be interpreted as comparing a shift in a function [namely, ${f(\epsilon) \mapsto f(\epsilon + \epsilon\alpha)}$, where $f(\epsilon) = T_n(1+\epsilon)-1$] to its first-order Taylor-series approximation. When this is a good approximation, anonymity is high, and little identifying information leaks out.

\begin{figure}[t!]
\includegraphics[width=\columnwidth]{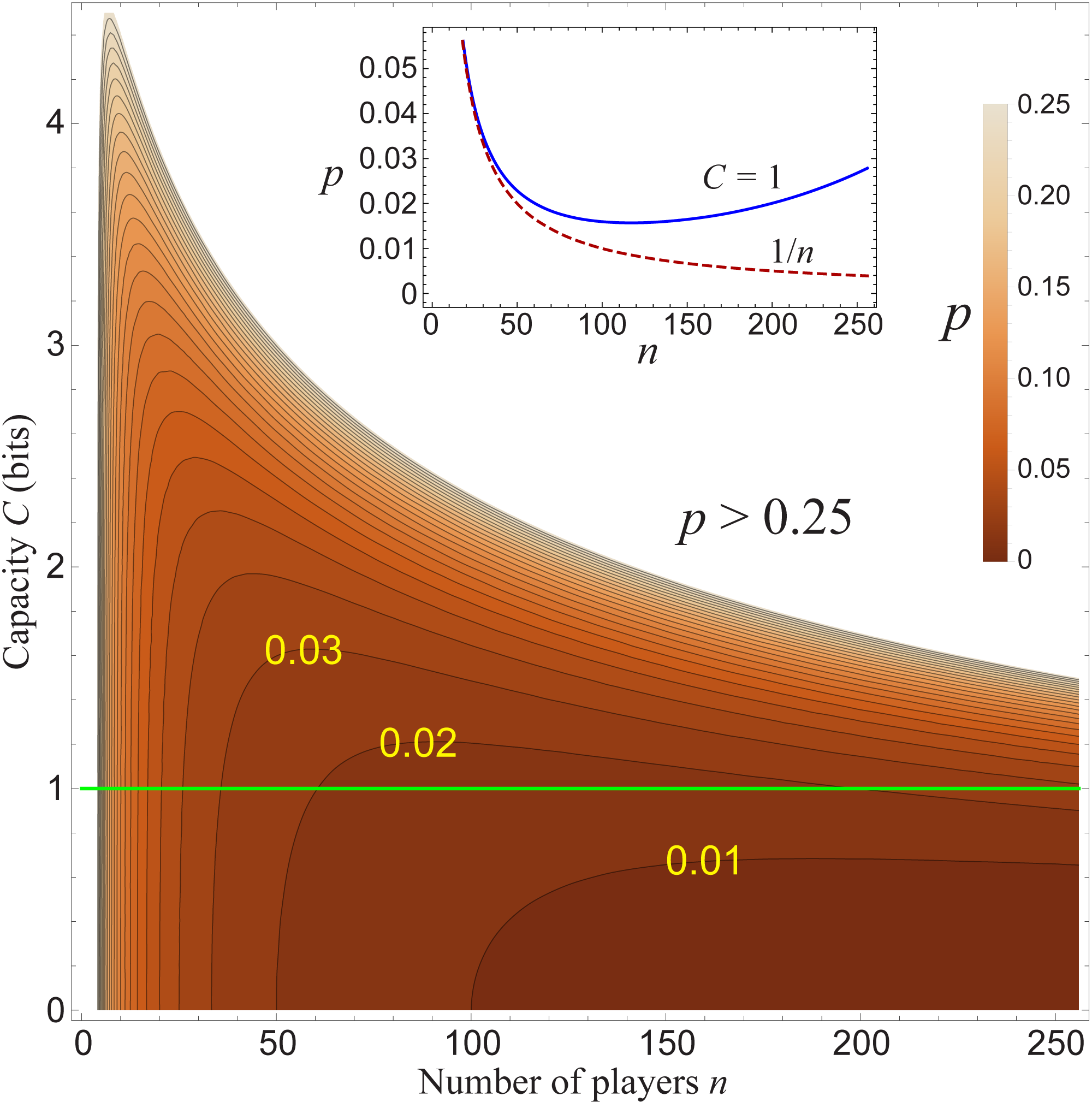}
\caption{\label{fig:results}Contour plot of the  geometric-mean probability~$p$ that the broadcaster is correctly identified during the protocol as a function of the number of players~$n$ and the channel capacity~$C$ [Eq.~\eqref{eq:channelcapacity}] in the limit of a large number of independent broadcast events. More precisely, we plot an upper bound on~$p$, which we calculate using Eq.~(\ref{eq:geomeanp}) and the upper bound for~$I({\bf M},A)$ from Eq.~\eqref{eq:mutinfobound}. %
Contours corresponding to $p=0.01,0.02,0.03$ are labeled, and subsequent contours increase by $0.01$ each. The white region corresponds to $p>0.25$. The squeezing is $20$~dB ($s = 10^{\text{(\#dB)}/20} = 10$), and each player's wedge width is $w=6$ (see Appendix~\ref{sec:width}). The inset shows (i)~a solid blue curve corresponding to a cross section of the main plot along the green $C=1$ line and (ii)~a dashed red curve corresponding to $p=1/n$. The latter corresponds to perfect tracelessness (no more risk than guessing randomly), which is only achieved in the trivial limit of no broadcast ($C=0$) or, for any $C>0$, in the asymptotic limit of infinite squeezing.}
\end{figure}

The only reason Eq.~\eqref{eq:mutinfobound} is not an equality is that we used the fact that the entropy of a mixture of Gaussians is upper bounded by the entropy of a Gaussian with the same covariance as that of the mixture. When this is a bad approximation, it is possible that the right-hand side of Eq.~\eqref{eq:mutinfobound} could exceed $H(A) = \log n$, while the actual value of $I(\vec M;A)$ never will.
Also note that $I(\vec M;A)$ as calculated is not additive under multiple repetitions of the protocol with the same broadcaster because after each run, the prior~$p_A(a)$ about the sender's identity will have changed based on the new information, requiring a new calculation. Nevertheless, 
 Eq.~\eqref{eq:mutinfobound} provides an estimate of the anonymity of the broadcaster in an asymptotic sense to be described shortly. (A calculation of single-shot probability of detection in a special case of the protocol is deferred to Sec.~\ref{sec:singleshot}.)

Anonymity is high whenever Alice's post-broadcast probability of discovery is very low. Since we have formulated the problem as a classical channel leaking (Shannon) information about Alice's identity, the relevant metric is the asymptotic behavior of the channel under $N$ independent uses for large~$N$, each with a (potentially) different broadcaster each time~\cite{Cover:2012ub}. Assuming each use is independent and the broadcaster and message are identically distributed each time, the asymptotic equipartition theorem states that the probability of a sequence~$(a_1, \dotsc, a_N)$ of broadcasters given %
$N$~independent broadcast events satisfies
\begin{align}
	\Pr(a_1, \dotsc, a_N)
	\approx
	2^{-N H(A | \vec M)}
\end{align}
with high probability~\cite{Cover:2012ub}. We can now define
\begin{align}
\label{eq:geomeanp}
	p
	\coloneqq
	2^{-H(A | \vec M)}
	= \frac {2^{I(\vec M;A)}} {2^{H(A)}}	
\end{align}
using log base 2%
. Since $p = \lim_{N\to\infty} \bigl[ \Pr(a_1, \dotsc, a_N) \bigr]^{1/N}$ with high probability, we can interpret $p$~as the \emph{geometric-mean probability that the broadcaster is correctly identified} over many independent broadcast events.

Note, however, that in any particular instance of the broadcast, $p$~would not be a valid estimate of the probability that that particular broadcaster is correctly guessed. We press on nonetheless using~$p$ because the analytic form of the mutual information makes it convenient for analysis. We perform the single-shot analysis using the (less accessible but more appropriate) min-entropy in Sec.~\ref{sec:singleshot}.

We want the quantity~$p$ to be small ($ p \ll 1$).
Replacing $I(\vec M;A)$ in Eq.~\eqref{eq:geomeanp} \blk with its upper bound from Eq.~\eqref{eq:mutinfobound} and then squaring both sides only strengthens the condition, which lets us write the following in the limit of a good resource state ($\epsilon \ll 1$):
\begin{align}
	n^2 
	&\gg 
	1+ \frac {(n^2-1)\alpha^2 \epsilon} {6(1+\alpha)} + O(\epsilon^2)\,.
\end{align}
Solving for $\alpha$ and dropping terms of $O(\epsilon^2)$ gives the bound:
\begin{align}
	\alpha \epsilon &\ll 6\,.
\end{align}
Since $\alpha \epsilon$ %
is the SNR of the broadcast message with respect to the excess noise in each of the local measurements, we can summarize this condition by saying that: \emph{anonymity is high when the broadcast message is sufficiently obscured by the local measurement noise}.

Clearly, there is a tradeoff between anonymity and channel capacity.\footnote{The variance restriction on the capacity is henceforth understood.} In particular, for a fixed value of the squeezing $s$, high SNR provides %
a larger channel capacity at the expense of lower anonymity. The opposite is also true:~small SNR corresponds to higher anonymity but smaller channel capacity. We explore this tradeoff in ~Fig.~\ref{fig:results} for a fixed squeezing  factor~${s = 10}$, corresponding to 20~dB. 
\section{Single-shot probability of detection} \label{sec:singleshot}

Here we consider a different scenario in order to make a more precise calculation of the guarantee of anonymity in a single-shot setting. As mentioned above, the min-entropy (rather than the Shannon entropy) is required for this. Furthermore, rather than considering channel capacity, which like the Shannon entropy is another asymptotic concept~\cite{Cover:2012ub}, here we consider a binary broadcast message rather than a real number being broadcast. Thus, our measure of success of the broadcast is in terms of the probability that the message is received as the opposite of what was sent (bit-flip error), and the probability of correct detection is calculated exactly in this single-shot scenario.

As before, player~$A$ (the broadcaster) is picked uniformly randomly from all players, but now players agree ahead of time on a simple binary encoding: \emph{only the sign of the broadcast message matters}. For simplicity, we will restrict to just two possible values of the real-valued broadcast message~$R \in \{ + r_0, - r_0 \}$. The probability that the binary message is received correctly is just the probability that~$M$ (the received broadcast message) has the same sign as~$R$ (the message being broadcast). For a particular message~$r \in \reals$, we saw in Sec.~\ref{subsec:abprotocol} that $M_{|r} \sim N_{r,(\Delta M)^2}(m)$, where $(\Delta M)^2 = \tfrac {1} {2s^2}$. By symmetry, for either choice of $r = \pm r_0$, the probability of misidentifying the binary broadcast message is therefore
\begin{align}
	p_{\text{err}}
= 
	\int_{-\infty}^{0} dm\, N_{r_0,(2s^2)^{-1}}(m)
=
	\frac 1 2 \erfc(s  r_0)
,
\end{align}
where the complementary error function $\erfc x = 1 - \erf x$. We can rearrange this to obtain the value of~$ r_0$ that gives a desired~$p_{\text{err}}$:
\begin{align}
\label{eq:r0fromperr}
	 r_0 = \frac 1 s \erfc^{-1} (2p_{\text{err}}).
\end{align}
Without loss of generality, we can assume that this (positive) value~$ r_0$ is the broadcast message since symmetry guarantees that the probability of discovery will not depend on the sign of the broadcast message, only on its magnitude.

We will need the following explicit definitions and probability calculations (see Appendix~\ref{sec:math} for notation):
\begin{align}
	A
\sim
	p_A(a)
&
=
	\frac 1 n
	,
\\
	\vec M_{|A}
\sim
	p_{\vec M|A}(\vec m|a)
&
=
	N_{r_0 \vec e_a, \mat \Sigma}(\vec m)
	,
\\
	(\vec M, A)
\sim
	p_{\vec M, A}(\vec m, a)
&
=
	\frac 1 n N_{r_0 \vec e_a, \mat \Sigma}(\vec m)
	,
\\
	\vec M
\sim
	p_{\vec M}(\vec m)
&
=
	\frac 1 n \sum_{a} N_{r_0 \vec e_a, \mat \Sigma}(\vec m)
	,
\\
	A_{|\vec M}
\sim
	p_{A|\vec M}(a | \vec m)
&
=
	\frac {N_{r_0 \vec e_a, \mat \Sigma}(\vec m)}
	{\sum_{a'} N_{r_0 \vec e_{a'}, \mat \Sigma}(\vec m)}
	,
\end{align}
where $\vec e_a$ is a vector of all zeros except for a~1 in slot~$a$. The distributions for~$A$ and~$\vec M_{|A}$ are prescribed, from which all of the others can be obtained using the laws of probability.
\blk

\begin{figure*}[t!]
\includegraphics[width=0.5\textwidth]{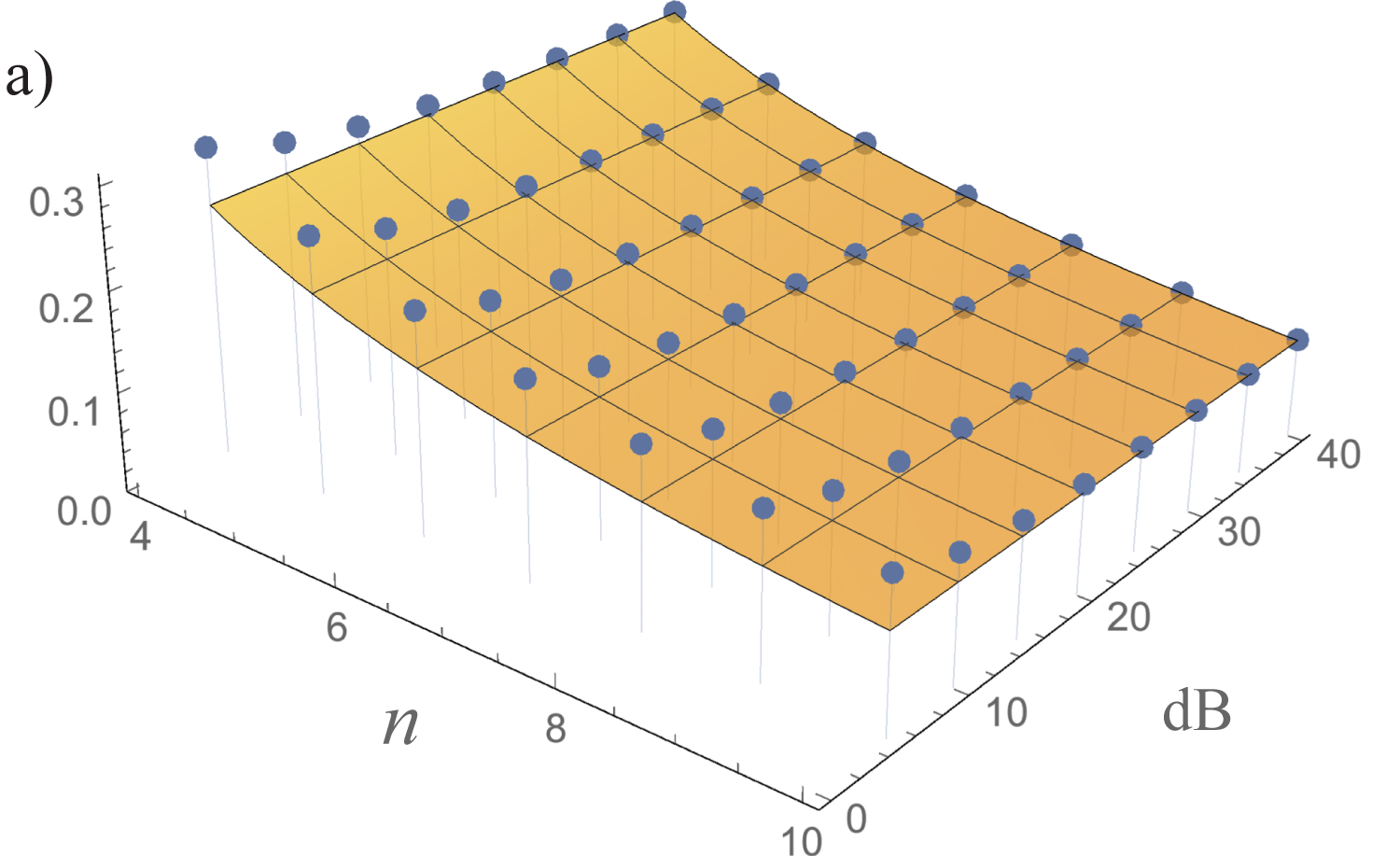}\hfill
\includegraphics[width=0.5\textwidth]{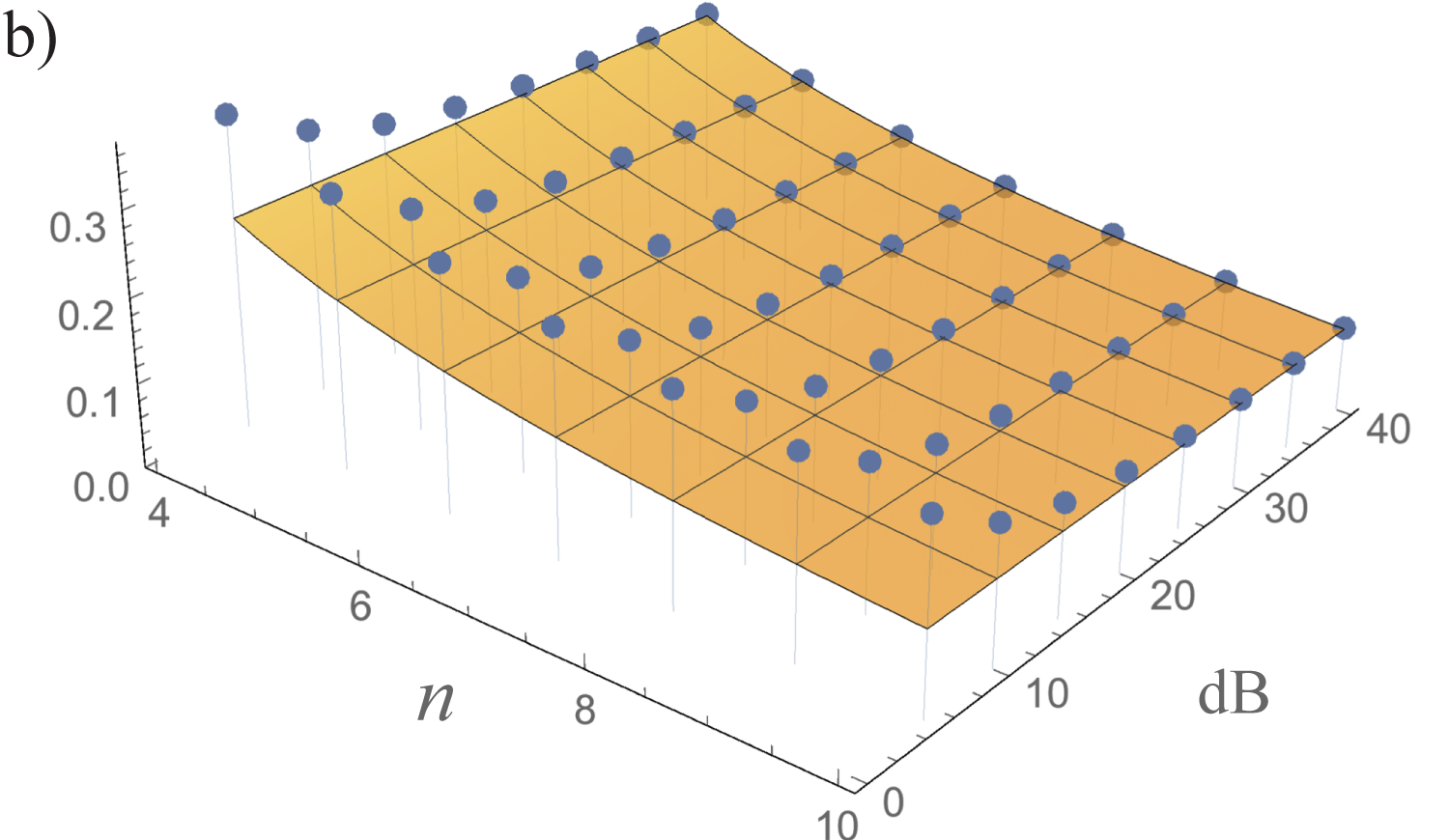}
\caption{\label{fig:singleshot} {  Average probability that the broadcaster is correctly identified in a single-shot broadcast of exactly one bit of data (i.e., only the sign of the broadcast message is recorded). The dots are plots of the probability to guess correctly, Eq.~\eqref{eq:pgAM}, for various values of~$n$ and amounts of squeezing (in~dB) with a wedge width $w=6$.  The orange plane is~$1/n$, which corresponds to uniformly random guessing. The broadcast magnitude~$r_0$ is scaled according to Eq.~\eqref{eq:r0fromperr} to correspond to 
(a) 1\% probability of bit-flip error and (b) 0.0001\%~probability of bit-flip error.} }
\end{figure*}

The min-entropy of a random variable~$X \sim p_X(x)$ is
\begin{align}
\label{eq:Hmin}
	H_{\text{min}}
\coloneqq
	-\log \max_x p_X(x)
\end{align}
The min-entropy (with log base~2) is related to the probability of guessing a random variable~$X$~\cite{Berens2013}. When given \emph{a particular set} of data~$y$, we can immediately write
\begin{align}
	p_{\text{g}}(X|{Y=y})
&
\coloneqq
	\Pr\bigl(\text{guess~$X$ correctly}|{Y = y} \bigr)
\nonumber \\
&
=
	2^{-H_{\text{min}}(X|{Y=y})}
\nonumber \\
&
=
	\max_x p_{X|Y}(x|y)
	.
\end{align}
To achieve this, one simply guesses that
\begin{align}
	X
&
=
	\arg \max_x p_{X|Y}(x|y).
\end{align}
That is, the best guess for $X$ is the highest-probability outcome~$x$ consistent with the data~$y$. Averaging over the data, one obtains the \emph{average} correct guessing probability~\cite{Berens2013}:
\begin{align}
	p_{\text{g}}(X|Y)
&
=
	\sum_y p_Y(y) p_{\text{g}}(X|{Y=y})
\nonumber \\
&
=
	\sum_y p_Y(y) \max_x p_{X|Y}(x|y)
	.
\end{align}

In our case,
\begin{align}
	p_{\text{g}}(A|\vec M)
&
=
	\int d^n m\, p_{\vec M}(\vec m) \max_a p_{A|\vec M}(a|\vec m)
\nonumber \\
&
=
	\frac 1 n \int d^n m\, \max_a N_{r_0 \vec e_a, \mat \Sigma}(\vec m)
	.
\end{align}
Simplifying this expression is somewhat involved. We start by noting the identity
\begin{align}
\label{eq:Gaussianidentity}
	N_{\vec 0, \mat \Sigma}(\vec x)
=
	\abs{\det \mat L} N_{\vec 0, \mat L \mat \Sigma \mat L^\tp}(\mat L \vec x)
\end{align}
for any invertible matrix~$\mat L \in \reals^{n \times n}$. Thus,
\begin{align}
&
	p_{\text{g}}(A|\vec M)
\nonumber \\
&
=
	\frac 1 n \int d^n m\, \max_a N_{\vec 0, \mat \Sigma}(\vec m - r_0 \vec e_a)
\nonumber \\
&
=
	\frac 1 n \int d^n m \, \abs{\det \mat \Sigma^{-1}} \max_a N_{\vec 0, \mat \Sigma^{-1}}(\mat \Sigma^{-1} \vec m - r_0 \mat \Sigma^{-1} \vec e_a).
\end{align}
Changing variables,
\begin{align}
	\vec u &= \mat \Sigma^{-1} \vec m,
\\
	 d^n u &= \abs{\det \mat \Sigma^{-1}} d^n m
	 ,
\end{align}
we have
\begin{align}
&
	p_{\text{g}}(A|\vec M)
\nonumber \\
&
=
	\frac 1 n \int d^n u \, \max_a N_{\vec 0, \mat \Sigma^{-1}}(\vec u - r_0 \mat \Sigma^{-1} \vec e_a)
\nonumber \\
&
=
	\frac {(\det 2\pi \mat \Sigma^{-1})^{-1/2}} n 
\nonumber \\
&
\qquad
\times 
	\int d^n u\, \max_a
	\exp
	\left[-\frac 1 2
	(\vec u - r_0 \mat \Sigma^{-1} \vec e_a)^\tp
	\mat \Sigma
	(\vec u - r_0 \mat \Sigma^{-1} \vec e_a)
	\right]
\nonumber \\
&
=
	\frac {(\det 2\pi \mat \Sigma^{-1})^{-1/2}} n
\nonumber \\
&
\qquad
\times 
	\int d^n u\, \max_a
	\exp
	\left[
	-\frac 1 2 \vec u^\tp \mat \Sigma \vec u
	+r_0 \vec u^\tp \vec e_a
	-\frac {r_0^2} {2} \vec e_a^\tp \mat \Sigma^{-1} \vec e_a
	\right]
\nonumber \\
&
=
	\frac {c_{\mat \Sigma}} {n}
	\int d^n u \, 
	N_{\vec 0,\mat \Sigma^{-1}}(\vec u)
	\max_a
	\exp(r_0 u_a)
\nonumber \\
&
=
	\frac {c_{\mat \Sigma}} n
	\int d^n u \, 
	N_{\vec 0,\mat \Sigma^{-1}}(\vec u)
	\exp\left( r_0 \max_a u_a \right)
	,
\label{eq:pgAMderiv}
\end{align}
where we have defined
\begin{align}
	c_{\mat \Sigma}
&
\coloneqq
	\exp
	\left(
		-\frac {r_0^2} {2} \vec e_a^\tp \mat \Sigma^{-1} \vec e_a
	\right)
=
	\exp
	\left(
		-\frac {r_0^2} {2n} \tr \mat \Sigma^{-1}
	\right)
\end{align}
by the fact that~$\mat \Sigma$ is invariant under permutation of the players' labels~$(a \to a+1)$.

Now we employ a trick: we carve up~$\reals^n$ into $n$ cones~$\{K_j\}_{j=1}^n$,
defined by
\begin{align}
	K_j \coloneqq \{\vec u \in \reals^n \mid u_k \leq u_j\; \forall k \neq j\}.
\end{align}
Intuitively, this is easy to understand: Every point~$\vec u \in \reals^n$ is an $n$-tuple of real numbers. The index~$j$ of the maximum entry of this $n$-tuple tells you which cone~$K_j$ the point belongs to. (In the case where there is more than one maximum entry, just choose the one with smallest index.) In this way, we can \emph{uniquely} partition~$\reals^n$ into these~$n$ cones---i.e., $\reals^n = \bigcup_{j=1}^n K_j$ (with any overlap of the cones being of measure~0). Thus we can write,
\begin{align}
\label{eq:pgAM}
	p_{\text{g}}(A|\vec M)
&
=
	\frac {c_{\mat \Sigma}} n
	\sum_{j=1}^n
	\int_{K_j} d^n u \, 
	N_{\vec 0,\mat \Sigma^{-1}}(\vec u)
	\exp\left( r_0 \max_a u_a \right)
\nonumber \\
&
=
	\frac {c_{\mat \Sigma}} n
	\sum_{j=1}^n
	\int_{K_j} d^n u \, 
	N_{\vec 0,\mat \Sigma^{-1}}(\vec u)
	\exp( r_0 u_j )
\nonumber \\
&
=
	c_{\mat \Sigma}
	\int_{K_1} d^n u \, 
	N_{\vec 0,\mat \Sigma^{-1}}(\vec u)
	\exp( r_0 u_1)
	,
\end{align}
where we used the fact that the value of the integral is the same for each cone~$K_j$. Notice that we never need to explicitly calculate~$\mat \Sigma^{-1}$. The final Gaussian has~$\mat \Sigma$ in the actual exponential, and $\tfrac 1 n \tr \mat \Sigma^{-1}$ (found within~$c_{\mat \Sigma}$) is just the harmonic mean of the eigenvalues of~$\mat \Sigma$.

We succeeded in partially analytically evaluating this integral, obtaining an expression that can be written solely in terms of the cumulative distribution function~(CDF) of a multivariate Gaussian. Unfortunately, it appears that there is no known analytic form for the CDF of a high-dimensional multivariate Gaussian. While various numerical techniques and approximations exist~\cite{Genz2009}, we found it sufficient for small~$n$ to have \emph{Mathematica} evaluate the integral as in Eq.~\eqref{eq:pgAM}. The results are shown in Fig.~\ref{fig:singleshot} for $p_{\text{err}} = 1\%$ and $p_{\text{err}} = 0.0001\%$ with several values of~$n$ and various levels of squeezing. 

The most important thing to note from these plots is that for large squeezing, the probability of correctly guessing the broadcaster $\sim 1/n$, which is no better than guessing randomly. Also note that for low squeezing, requiring a lower $p_{\text{err}}$ (the chance of a bit flip in the message) increases the risk that the broadcaster is correctly identified. This is consistent with the tradeoff we found between channel capacity and anonymity in the asymptotic analysis of Sec.~\ref{sec:anonymity}. Ideally, we would like to be able to see whether the same phenomenon appears for large~$n$ in this case that we found in the asymptotic analysis---i.e., increased risk of detection for large~$n$. Due to numerical limitations, we were unable to evaluate this case for large~$n$, so we leave this as an open question.
\blk

\section{Error mitigation by reservoir engineering} \label{sec:error_mit}

After preparation, a CV toric-code state can be protected from errors arising from decoherence and other sources while the players await the broadcasting protocol.  Here we present a proof-of-principle calculation to illustrate the method; we leave a full derivation to a future publication. 

For simplicity, we focus on creating the toric-code logical vacuum state~$\ket \GSvac$, Eq.~\eqref{eq:GSvac}. Note that this is not the same as the state used in the analysis of the broadcasting protocol above---that being the toric-code logical squeezed state~$\ket \GSs$, Eq.~\eqref{eq:GSvac}. We choose the vacuum state however because it most clearly illustrates the basics of the method, which
relies on reservoir engineering \cite{Diehl:2008ha}, where a dynamical master equation typically drives the system towards a desired steady state.

This is achieved by coupling the physical modes to bosonic reservoirs, $\{\op b_i(\omega)\}$, at each vertex and face of the lattice; see Fig.~\ref{fig:TCandprotocol}(d). The mode-reservoir coupling is described by a quadratic, quasi-local Hamiltonian 
\begin{align}
	\op H_{\text{int}} =  \sum_{i \in \{\mathcal{V},\mathcal{F}\}} \int d\omega \,  \kappa(\omega) \left[\hat{\eta}_i \op b_i^\dagger(\omega) +  \hat{\eta}^\dagger_i \op b_i(\omega) \right],
\end{align}
 where $[\op b_i(\omega), \op b_j^\dagger(\omega')] = \delta_{i,j} \delta(\omega - \omega')$. Tracing out the reservoirs in the usual Markov and rotating-wave approximations yields a map in Lindblad form with the CV toric-code nullifiers~$\op \eta_i$ as jump operators, 
	\begin{align} \label{eq:mitigatemap}
		\mathcal{L}_\mitigate[ \op{\rho}] = \sum_{i \in \{ \mathcal{V}, \mathcal{F}\} } 
		\left( \op{\eta}_i \op{\rho} \op{\eta}_i^\dagger - \frac{1}{2} \big\{ \op{\eta}_i^\dagger \op{\eta}_i, \op{\rho} \big\}_+ \right),
	\end{align}
and decay rate $\gamma_\mitigate = 2 \pi |\kappa(\omega_0)|^2$ arising from evaluation of the coupling strength at frequency $\omega_0$ \cite{GarZol04}. For finite squeezing, the nullifiers in Eqs.~(\ref{eq:nullifiers}) are not Hermitian, and the map in Eq.~(\ref{eq:mitigatemap}) cools by extracting entropy from the Hilbert space spanned by the nullifiers. The map in Eq.~(\ref{eq:mitigatemap}) drives the state towards the codespace (i.e.,~the nullspace of the nullifiers),  $\hat{\rho} \rightarrow \op{\rho}_
\GS = \ket{\bs{\eta}}\bra{ \bs{\eta} }_{\rm null} \otimes \op{\rho}_\logic$, where $\op{\rho}_\logic$ is in general a mixed state in the logical modes that depends on the initial state.  

\begin{figure}[t]
\includegraphics[width=\columnwidth]{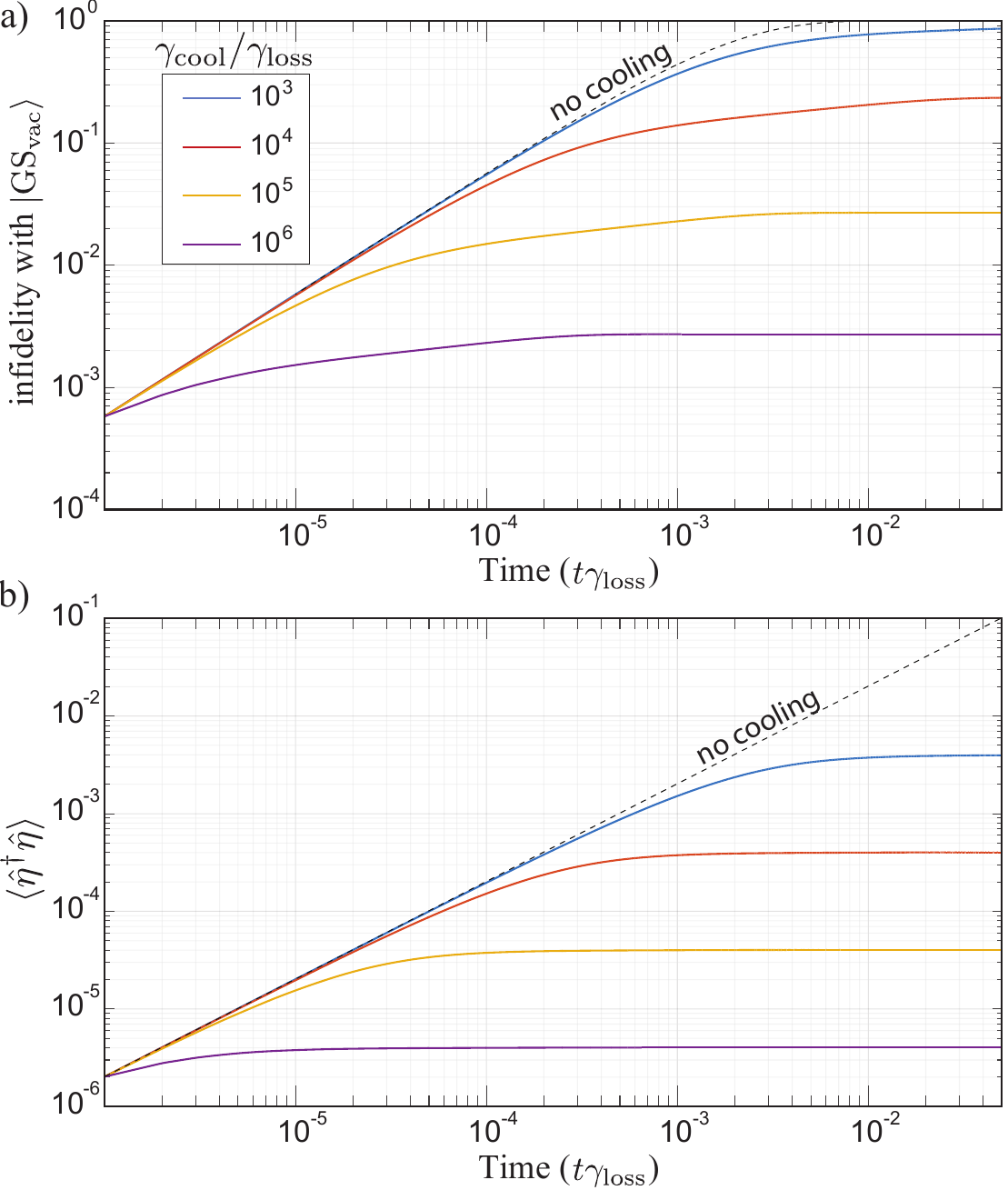}
\caption{\label{fig:mitigate}  Example of state maintenance via dissipative error mitigation. 
The initial state is the CV toric-code vacuum $\ket \GSvac$, Eq.~\eqref{eq:GSvac}, defined by the symmetric nullifiers, Eqs.~\eqref{eq:nullifiers}, on a $6 \times 24$ lattice with 10~dB of squeezing (${s = \sqrt{10}}$). (a)~Infidelity with $\ket \GSvac$, $1 - \mathcal{F}(\hat{\rho},\hat{\sigma})$, where fidelity is given by Eq.~(\ref{eq:fidelity}). %
(b)~Excitation number, $\mean{ \op{\eta}^\dagger \op{\eta} }$, for a single nullifier (identical for face or vertex). In both (a) and (b) the curves are ordered $\gamma_{\rm cool}/\gamma_{\rm loss} = \{ 10^3, 10^4, 10^5, 10^6\}$} from top to bottom.
\end{figure}

During maintenance of a CV toric code, the cooling provided by Eq.~(\ref{eq:mitigatemap}) competes against errors.  A local error, e.g.,~photon loss on a single mode, takes the system outside the nullspace of some or all of the nullifiers touching that mode. The map in Eq.~(\ref{eq:mitigatemap}) returns the state to the codespace at the expense of mixedness within the logical modes. To illustrate performance, we assume local photon loss with a uniform rate for all modes, although such cooling can be effective against more general errors including those that are asymmetric, nonlocal, and correlated. 

Here, we consider the evolution of the state of the collection of modes, $\hat{\rho}$, under the cooling in Eq.~(\ref{eq:mitigatemap}) while each physical mode is subject to photon loss at rate $\gamma_{\rm loss}$. These dynamics are described by the mast\'{e}r equation
	\begin{align} \label{eq:masterequation}
		\frac{d}{dt} \op{\rho} = \gamma_\mitigate \mathcal{L}_\mitigate[ \op{\rho}] + \gamma_{\rm loss} \sum_{e\in \mathcal{E}} \left( \op a_e \op{\rho} \op a_e^\dagger - \frac{1}{2} \big\{ \op a_e^\dagger \op a_e, \op{\rho} \big\}_+ \right).
	\end{align}
The cooling map in Eq.~\eqref{eq:mitigatemap}, which is implemented quasi-locally, damps out unwanted errors. Since such error protection is not active error correction, we refer to it as \emph{mitigation}.

In order to keep the focus of this work on the broadcasting protocol, we defer the details of this mitigation process to a separate publication%
. To illustrate the benefit of this method, however, we begin with a CV toric-code vacuum state~$\ket{\GSvac}$ from Eq.~\eqref{eq:GSvac} using the symmetric nullifiers from Eqs.~\eqref{eq:nullifiers}. This state then undergoes simultaneous local loss with rate~$\gamma_{\rm loss}$. Local loss leads to the state decaying to the local vacuum of all modes, but this process can be kept in check by error mitigation as shown in Fig. ~\ref{fig:mitigate}.

For Gaussian dynamics, the evolution can be described entirely by the quadrature means and covariance matrix as described in Appendix~\ref{sec:GDynamics}.
We quantify the performance of the error mitigation
by the Uhlmann-Jozsa fidelity~\cite{Jozs94},
	\begin{align} \label{eq:fidelity}
		\mathcal{F}(\hat{\rho}, \op{\sigma}) = \left[ \tr \bigg( \sqrt{ \sqrt{ \op{\rho} } \op{\sigma} \sqrt{\op{\rho}}  } \bigg) \right]^2.
	\end{align} 
For the pure target state $\op{\sigma} = \ket{\GSvac}\bra{\GSvac} $, the fidelity reduces to $\mathcal{F}(\op{\rho}, \op{\sigma}) = \tr (\op{\rho} \hat{\sigma})$, which can be evaluated directly from the covariance matrices using the formula $\mathcal{F}(\op{\rho}, \op{\sigma}) = [ \text{det} (\mat \Sigma + \mat \Sigma_\sigma) ]^{-1/2}$~\cite{fidelity13}. Figure~\ref{fig:mitigate}(a) shows the improved fidelity for increasing cooling rates, illustrating error mitigation. 

All CV toric-code states satisfy the condition that ${\op{\eta}_i\ket{ \GS }  = 0}$ for all nullifiers. A measure of the degree to which this condition is violated, and thus  the degree to which  the state leaves the codespace, is the nullifier excitation number~$\mean{ \op{\eta}^\dagger \op{\eta} }$. Fig.~\ref{fig:mitigate}(b) shows the protection of the codespace as the nullifier excitation number is stabilized by the cooling map, Eq. (\ref{eq:mitigatemap}). After a relaxation time that scales with the lattice size, the system approaches a steady state. For strong cooling, $\gamma_{\rm loss}/\gamma_\mitigate \ll 1$, one finds that the expectation value of the nullifier number operators reaches a steady-state~(ss) value that scales $\mean{ \op{\eta}_v^\dagger \op{\eta}_v }_{\rm ss} = \mean{ \op{\eta}_f^\dagger \op{\eta}_f }_{\rm ss} \propto \gamma_{\rm loss}/\gamma_\mitigate$%
. Thus, the steady state is close to the toric-code vacuum, $\op{\rho}_{\rm ss} \sim \ket{\GSvac}\bra{\GSvac}$.

\section{Optical implementation}
\label{sec:implementation}

This protocol may be implemented using recently demonstrated methods for generating large-scale optical CV cluster states encoded in either frequency modes~\cite{Wang:2014im,Chen:2014jx} or temporal modes~\cite{Menicucci2011a,Yokoyama:2013jp}. %
The GHZ-state version is achievable now with achieved squeezing levels (5~dB) in current technology~\cite{Yokoyama:2013jp}. Proof-of-principle experiments with a surface-code state are possible with $\sim$10~dB of squeezing, which is state of the art but achievable~\cite{Eberle:2010jh,Mehmet:2011bu,Yokoyama:2013jp}. Higher squeezing would enable practical large-scale anonymous broadcasting.%

Resource states could also be prepared in circuit-QED setups, either dynamically or by engineering a quadratic Hamiltonian between microwave cavities~\cite{PazSilva:2009fa} that has the CV cluster state as the gapped ground state and then performing quadrature measurements to map it to a CV surface code~\cite{Demarie:2014jx}. Single-mode~\cite{Yurke:1988ih,CastellanosBeltran:2008cg} and two-mode~\cite{Flurin:2012hq,Bergeal:2012hj,Eichler:2011fc,Wilson:2011ir} squeezing has already been demonstrated in these systems, and the SQUID-based controlling technology allows for very strong nonlinearities~\cite{Devoret:2007wo,Ashhab:2010eh,Allman:2014dn}, enabling high squeezing ($\sim$13~dB)~\cite{Moon:2005jr,Zagoskin:2008kp,YiHuo:2008ke,Li:2011he,Grimsmo:2017aa}.
	\subsection{Macrocode-based CV cluster states}

Recent experimental results have shown that compact optical experimental setups can produce huge CV cluster states, including million-mode~\cite{Yoshikawa2016} and $10^4$-mode CV cluster states~\cite{Yokoyama:2013jp} with modes multiplexed in time (temporal modes) and a 60-mode CV cluster state~\cite{Chen:2014jx} with modes multiplexed in frequency (frequency modes). These are cluster states with linear graphs, but the extension to a square lattice is straightforward and readily achievable with current technology~\cite{Menicucci2011,Wang:2014im,Menicucci2008}.

These setups were already discussed in Ref.~\cite{Demarie:2014jx} as candidates for generating CV surface-code states like the ones necessary for this protocol. Here we review this construction and discuss its implementation for anonymous broadcasting.

\begin{figure}
\begin{align*}
	\mat G = \raisebox{-0.5\height+0.75ex}{\includegraphics[width=.70\columnwidth]{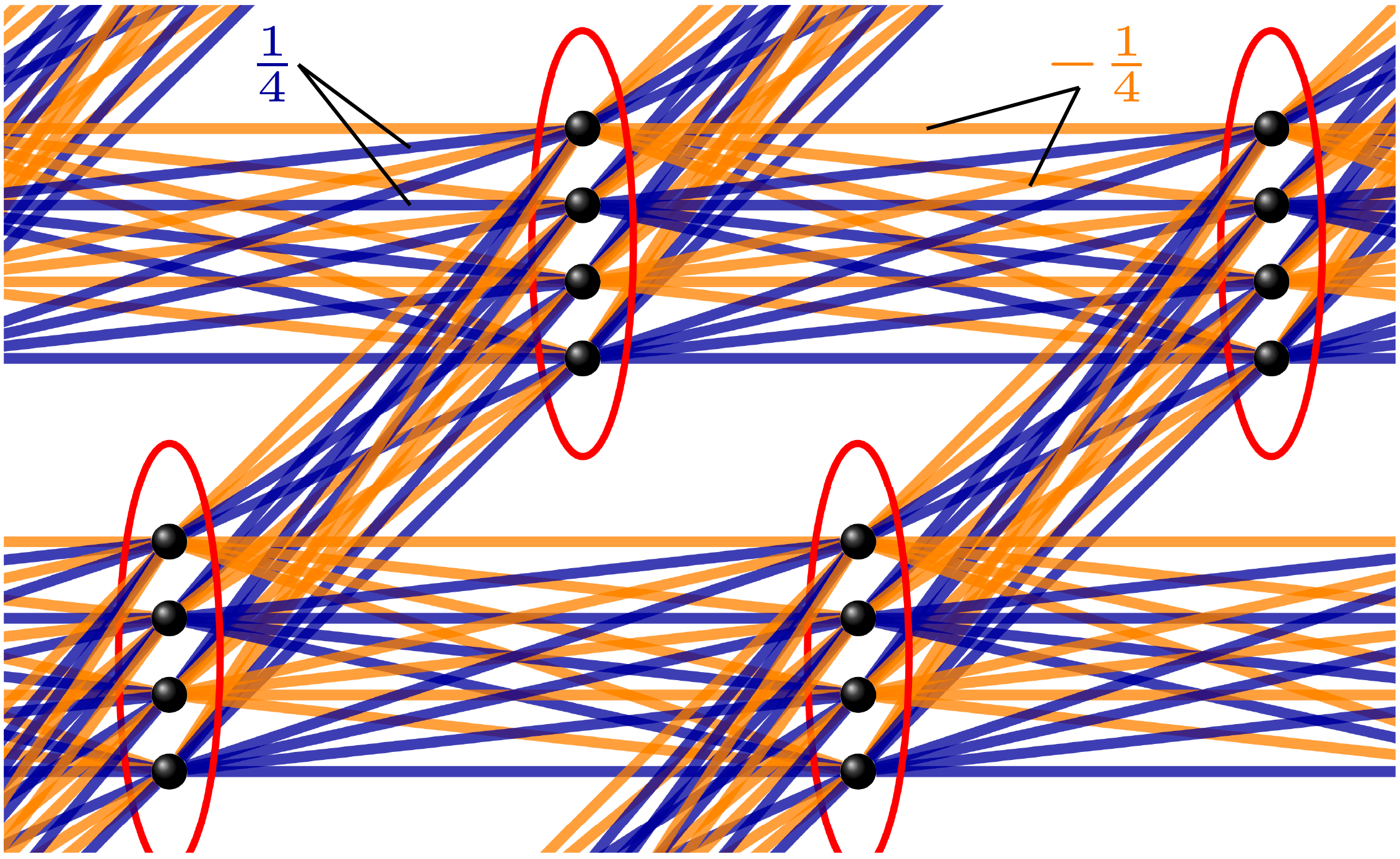}}
\end{align*}
\caption{\label{sup:fig:macronodes}Basic graph~$\mat G$ for temporal-mode CV cluster states~\cite{Menicucci2011a}; the full graph~\cite{Menicucci2011} is given in Eq.~\eqref{eq:Zoptics}. $\mat G$, as shown, also represents frequency-mode CV cluster states~\cite{Menicucci2008,Flammia2009,Wang:2014im} up to trivial $\pi$ phase shifts that merely flip the sign of some of the edges. Notice that $\mat G$ has the overall structure of a square lattice%
, but the individual nodes of that lattice are now collections of 4 nodes called \emph{macrocodes}. Each macronode is identified by its surrounding red oval. In the temporal-mode case~\cite{Menicucci2011a}, each of the 4 nodes within a macronode is a synchronous temporal mode in four spatially separate laser beams. In the cylindrical frequency-mode case~\cite{Wang:2014im}, each of the 4 nodes within a macronode share a common frequency but differ in spatial beam and polarization. The toroidal frequency-mode case~\cite{Menicucci2008,Flammia2009} is more complicated in structure and offers no advantages over the cylindrical one, so we do not consider it further.}
\end{figure}

The temporal-mode~\cite{Menicucci2011a,Yokoyama:2013jp} and frequency-mode~\cite{Menicucci2008,Flammia2009,Wang:2014im,Chen:2014jx} construction methods generate a toroidal~\cite{Menicucci2008,Flammia2009} or cylindrical~\cite{Menicucci2011a,Wang:2014im} CV cluster state with a Gaussian graph~\cite{Menicucci2011} %
whose overall structure is that of a square lattice but is nevertheless not an ordinary  square lattice.  %
Instead, it is a lattice based on 4-node groupings called \emph{macronodes}, with a structure as shown in Fig.~\ref{sup:fig:macronodes}. The actual CV cluster state has the full graph~\cite{Menicucci2011}
\begin{align}
\label{eq:Zoptics}
	\mat Z = i\delta \mat I + t \mat G\,,
\end{align}
where $\delta = \sech 2\sqp$, $t = \tanh 2\sqp$, and $\mat G$ is the graph shown in Fig.~\ref{sup:fig:macronodes}, with edge weights $\pm \tfrac 1 4$.

By measuring the top three modes of each macronode in $\op q$, all but a single layer of the grid is deleted, leaving a uniformly-weighted, ordinary CV cluster state with graph~\cite{Menicucci2011}
\begin{align}
	\mat Z_{\text{CS}} = i\delta \mat I + g \mat A_{\text{grid}}\,,
\end{align}
where $\delta = \sech 2\sqp$, $g = \tfrac 1 4 \tanh 2\sqp$, $\sqp > 0$ is an overall squeezing parameter, and $\mat A_{\text{grid}}$ is a binary adjacency matrix for an ordinary square-lattice graph with boundary conditions (toroidal or cylindrical) inherited from its parent, Eq.~\eqref{eq:Zoptics}. Note that the edge weights in~$\mat Z_{\text{CS}}$ are all~$\tfrac 1 4 \tanh 2\sqp$, while in the canonical construction, %
they should all be~1. Nevertheless, we can remodel the cluster state~\cite{Alexander:2014ew,Demarie:2014jx} by redefining quadratures so that the edge weights are~1 but at a cost of multiplying the self-loop weights by~$g^{-1}$. Since $\sech 2\sqp = \delta \eqqcolon s_0^{-2}$, this means that the original value of $s_0$ (so labeled to differentiate it from the actual $s$ used in the protocol) could be considered to be $s_0 = \sqrt{\cosh 2\sqp}$, except for the non-unit~$g$. The new effective value of~$s$, which should be used in the calculations in the previous sections, is less than half this initial value~\cite{Alexander:2014ew,Demarie:2014jx}:
\begin{align}
\label{eq:swithalpha}
	s &= \frac {s_0} {2} \sqrt{\tanh 2\sqp} %
	= \frac 1 2 \sqrt{\sinh 2\sqp}\,,
\end{align}
With a canonical CV cluster state obtained, which has uniform edge weight of~1, with $s$ from Eq.~\eqref{eq:swithalpha}, we can use local $\op q$ measurements to ``cut and unroll'' the cylinder or torus into a square lattice with the necessary smooth/rough boundary conditions as identified in Appendix~\ref{subsec:open}. Further local $\op q$ and $\op p$ measurements are then used to convert this state to a CV surface code state~\cite{Demarie:2014jx} with two rough and two smooth edges as shown in Fig.~\ref{sup:fig:open}(b), which is then distributed to the players. The broadcast protocol proceeds according to the modifications described in Appendix~\ref{subsec:open}.

One might think we could take advantage of the cylindrical or toroidal structure of the original CV cluster states to produce a surface-code state with periodic boundaries. This fails, however, because the graphs of both states have a one-grid-unit twist along each compactified direction~\cite{Wang:2014im,Menicucci2011a,Flammia2009}, which makes the checkerboard pattern of measurements needed to convert it into a cylindrical or toroidal surface code fail to line up properly. This is why we have to cut it into a surface code with open boundaries instead. If the twist were by an even number of grid units, other boundary conditions might be possible.

The temporal-mode scheme~\cite{Menicucci2011a} claims an advantage over the cylindrical frequency-mode scheme~\cite{Wang:2014im} in terms of ease of distribution. This is because the temporal-mode cylindrical lattice is built up like sequentially winding thread around a spool. This means that large chunks of the lattice are contiguous in time. Thus, one only needs a quickly adjustable mirror in order to distribute the pieces of the lattice to the players. Initially, the mirror is used to direct one of the four output beams to the first player. (The other three beams are immediately measured in $\op q$ to do the projection down to an ordinary lattice.) Once the player has received enough modes to form his/her sublattice, the mirror is switched so that the output beam is directed toward the second player, and so on. $\op q$ measurements at the start and end of this entire process are used to clean up the total lattice before the players themselves do the necessary additional $\op q$ and $\op p$ measurements to transform the state into a surface-code state. The ``radius of the cylinder'' in the temporal-mode case is limited by the coherence length~$L$ of the laser, but its width in the temporal direction---which is the direction used to measure the width~$w$ of each player's wedge, for instance---is not so limited since far-separated modes do not need to directly interact. This means that the temporal-mode scheme is capable of involving a practically unlimited number of players. \blk

The cylindrical frequency-mode scheme~\cite{Wang:2014im} has the same graph structure, but the frequencies of nearby modes are widely separated, so it is not as easy to split the lattice up into contiguous pieces for distribution. If this hurdle could be overcome, the frequency-mode scheme might claim an advantage because %
it is a continuous-wave scheme, meaning it might provide a means to transmit information continuously, rather than in bursts, as would be required by the temporal-mode scheme.

	\subsection{Squeezing levels for surface-code protocol}

The rescaling of $s$ shown in Eq.~\eqref{eq:swithalpha} means that this is likely not the most efficient way of generating a surface-code state, in terms of making good use of available squeezing resources~\cite{Alexander:2014ew}. Further theoretical work could lead to better procedures, but for now, we can look at the state of the art and what is achievable.

The largest squeezing achieved to date in these large-scale schemes is 5~dB %
in the temporal-mode experiment~\cite{Yokoyama:2013jp}. This corresponds to \footnote{Here and throughout, the abbreviation ``$\#$dB" stands for ``number of decibels."}
\begin{align}
\label{eq:sqpfromdB}
	\sqp = \frac {\#\text{dB}} {20} \ln 10 \simeq 0.5756\,,
\end{align}
which means that the effective $s$ for a protocol using this state is
\begin{align}
\label{eq:sfromsqp}
	s = \frac 1 2 \sqrt{\sinh 2\sqp} \simeq 0.5965\,,
\end{align}
which corresponds to an effective initial squeezing %
of
\begin{align}
\label{eq:effdB}
	\text{(effective \#dB)} = 20 \log_{10} s \simeq -4.488~\text{dB}
\end{align}
when doing the protocol. The negative sign means that this state is equivalent to a canonical CV cluster state [Fig.~\ref{sup:fig:toric}(a)] made with \emph{anti-squeezed} vacuum modes (i.e.,~vacuum modes squeezed in the wrong direction)%
~\cite{Menicucci2011}%
. Note that this does not mean that we would be better off not doing any squeezing at all in the actual experiment. Instead, this is simply a side-effect of the straightforward, but squeezing-inefficient~\cite{Alexander:2014ew,Demarie:2014jx}, projection to an ordinary lattice from the macrocode-based lattice shown in Fig.~\ref{sup:fig:macronodes}. In this case, it produces a poor-quality state that is equivalent to one made with anti-squeezed input modes. Since we want $s^2 \gg 1$ for nontrivial channel capacity with high anonymity, either improved squeezing or further theoretical improvements in the protocol would be required to make practical use of these resources.

Single-mode squeezing as high as 12.7~dB~\cite{Eberle:2010jh,Mehmet:2011bu}, and even 15~dB~\cite{Vahlbruch2016}, has been achieved in optics experiments, 
so it would be state of the art, but not unreasonable, to consider 10~dB achievable in temporal-mode~\cite{Yokoyama:2013jp,Menicucci2011a} or frequency-mode~\cite{Chen:2014jx,Wang:2014im} CV cluster states. Using Eqs.~\eqref{eq:sqpfromdB}, \eqref{eq:sfromsqp}, and~\eqref{eq:effdB}, this corresponds to an effective squeezing of $+0.925$~dB, or an effective $s=1.112$. This would still allow for semi-anonymous broadcasting---which we define as giving a probability~$p<2/n$ of the sender being correctly identified (less than twice the probability of random guessing). This would be possible when broadcasting $0.25$~bits (corresponding to an SNR $\alpha = 0.414$) for $n\le11$ or broadcasting $0.5$~bits ($\alpha = 1$) with $n\le 5$. This would be enough for a proof-of-principle demonstration.

\subsection{Squeezing levels for GHZ-state protocol}

The calculations above assume that a full surface-code state is used as the resource. This has a macrocode-based graph with edge weights~$\pm \frac 1 4$, as shown in Fig.~\ref{sup:fig:macronodes}, which reduces the effective squeezing dramatically when projected down to an ordinary lattice~\cite{Alexander:2014ew}. A surface code is necessary for error mitigation but not for basic demonstration of the protocol itself. For this, a simple GHZ state will suffice. As shown in  Appendix~\ref{subsec:GHZ}%
, this can be made from a linear CV cluster state.

\begin{figure}[t]
\begin{align*}
	\mat G = \raisebox{-0.5\height+0.75ex}{\includegraphics[width=.70\columnwidth]{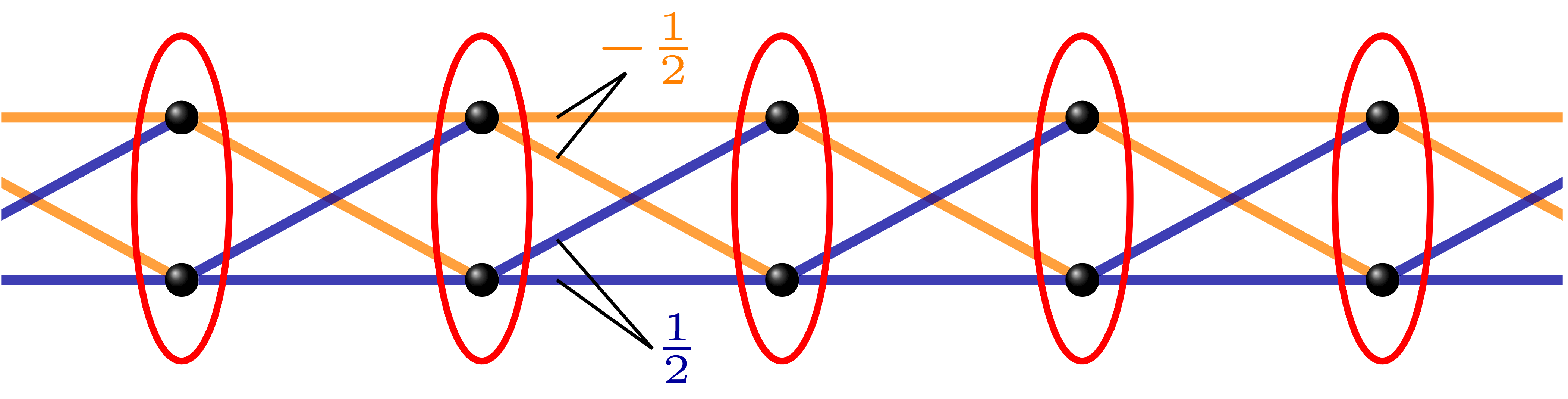}}
\end{align*}
\caption{\label{sup:fig:macrolinear}Basic graph~$\mat G$ for the temporal-mode linear CV cluster state reported in~\cite{Yokoyama:2013jp}; the full graph~$\mat Z$~\cite{Menicucci2011} is obtained from this through Eq.~\eqref{eq:Zoptics}. $\mat G$, as shown, also represents frequency-mode CV cluster states reported in~\cite{Chen:2014jx} up to trivial $\pi$ phase shifts that merely flip the sign of some of the edges. Notice that $\mat G$ has the overall structure of a line graph%
, but the individual nodes of that lattice are now collections of 2 nodes called \emph{macrocodes}. Each macronode is identified by its surrounding red oval. In the temporal-mode experiment~\cite{Yokoyama:2013jp,Menicucci2011a}, each node within a macronode is a synchronous temporal mode in spatially separate laser beams. In the frequency-mode experiment~\cite{Chen:2014jx,Wang:2014im}, each node is one of two polarizations with the same frequency.}
\end{figure}

The basic graph~$\mat G$ for the actual state created in the temporal-mode experiment~\cite{Yokoyama:2013jp} is shown in Fig.~\ref{sup:fig:macrolinear}, where the full graph~$\mat Z$~\cite{Menicucci2011} is again obtained from~$\mat G$ through Eq.~\eqref{eq:Zoptics}. This graph has 2-node macrocodes (instead of 4-node), and the edge weights are~$\pm \frac 1 2$ (instead of $\pm \frac 1 4$), which means that with a base squeezing of 5~dB, the effective~$s$ for a protocol based on this linear resource~\cite{Alexander:2014ew} is larger than in the surface-code case [compare Eq.~\eqref{eq:sfromsqp}]:
\begin{align}
\label{eq:sfromsqplinear}
	\sqp &\simeq 0.5756 &\Longrightarrow&&  s &= \frac {1} {\sqrt 2} \sqrt{\sinh 2\sqp} \simeq 1.006\,.
\end{align}
This corresponds to an effective initial squeezing of
\begin{align}
\label{eq:effdBlinear}
	\text{(effective \#dB)} = 20 \log_{10} s \simeq +0.05297~\text{dB}\,,
\end{align}
which can be compared with Eq.~\eqref{eq:effdB}.

With error correction not possible when using a GHZ state, we can reduce the wedge width~$w$ to its minimum value: $w=1$. In this scenario, semi-anonymous broadcasting ($p<2/n$; see subsection above) is possible for
\begin{align}
	C &= 0.25~\text{bits} & (\alpha &= 0.414)\,, & n&\le17\,; \\
	C &= 0.5~\text{bits} & (\alpha &= 1)\,, & n&\le8\,; \\
	C &= 0.75~\text{bits} & (\alpha &= 1.828)\,, & n&\le5\,; \\
	C &= 1~\text{bit} & (\alpha &= 3)\,, & n&\le4\,.
\end{align}
Thus, optical technology available today~\cite{Yokoyama:2013jp} can be used to demonstrate a practical implementation of GHZ-state-based anonymous broadcasting using this protocol.%

\subsection{Scalability}

The main advantage of these optical implementations remains in their immense scalability. CV GHZ states are already available today with current technology for anonymous broadcasting, and surface-code-based protocols are possible with state-of-the-art implementations. If the squeezing can be increased (or a more efficient conversion protocol devised), this technology holds great promise for large-scale anonymous broadcasting.

\section{Conclusion}

We propose using large-scale continuous-variable topological quantum codes for the important practical task of anonymously broadcasting classical information, and we quantify the channel capacity and anonymity of the protocol in terms of its physical parameters. Large squeezing enables high-capacity broadcasting with strong anonymity, but there is a trade-off between the two for any fixed level of squeezing. %
Our protocol outperforms other anonymous broadcasting protocols in two crucial ways: (1)~Because a topological quantum code serves as the resource, the scheme is robust to errors and further can be protected with quasi-local reservoir engineering. (2)~Because that code is a continuous-variable code, the technology required for large-scale resource generation is already available%
.  A notable feature of our protocol using continuous variables (instead of qubits) is that with large enough squeezing, anonymity is maintained even with channel capacity $C > 1$~bit. This would enable other, more complex tasks such as anonymous yes-or-no voting~\cite{BJT10,Jiang:2012ja} within a group of size~$\le C$. %

\section*{ACKNOWLEDGMENTS}%
G.K.B.\ thanks James Wootton for discussions on the discrete variable protocol and thanks J.~Dowling for comments. T.F.D.\ thanks Joseph Fitzsimons for useful suggestions about past work on anonymous communication. B.Q.B.\ thanks Naoki Yamamoto and Ian Petersen for valuable comments. N.C.M.\ thanks Marco Tomamichel for useful discussions. T.F.D.\ is supported by the Singapore National Research Foundation under NRF Award No. NRF-NRFF2013-01. N.C.M.\ is supported by the Australian Research Council~(ARC) under grant No. ~DE120102204 and by the U.S.\ Defense Advanced Research Projects Agency (DARPA) Quiness program under Grant No.\ W31P4Q-15-1-0004. N.C.M.\ and B.Q.B\ acknowledge support from the ARC Centre of Excellence for Quantum Computation and Communication Technology (Project No.\ CE170100012). G.K.B.,\ T.F.D.,\ and B.Q.B.\ acknowledge support from the ARC Centre of Excellence for Engineered Quantum Systems (Project No.\ CE110001013). G.K.B. acknowledges support from ARC Project No. DP160102426.
\bibliographystyle{bibstyleNCM_papers}
\bibliography{CVTAB}

\begin{thebibliography}{10}

\bibitem{Movahedi:2014tt}
M. Movahedi, J. Saia, and M. Zamani, ``Secure Anonymous Broadcast,''
  arXiv:1405.5326v1  (2014).

\bibitem{Stajano:2000kh}
F. Stajano and R. Anderson, ``{The Cocaine Auction Protocol: On the Power of
  Anonymous Broadcast},''  in {\em Information Hiding} (Springer Berlin
  Heidelberg, Berlin, Heidelberg, 2000), \ pp.\ 434--447.

\bibitem{T.-Ruffing:2016aa}
T. Ruffing, P. Moreno-Sanchez, and A. Kate, ``P2P Mixing and Unlinkable Bitcoin
  Transactions,'' Cryptology ePrint Archive Report 2016/824 (2016).

\bibitem{Chaum:1988dg}
D. Chaum, ``{The dining cryptographers problem: Unconditional sender and
  recipient untraceability},'' J. Cryptology {\bf 1}, 65 (1988).

\bibitem{BT07}
A. Broadbent and A. Tapp, ``Information-theoretic security without an honest
  majority,'' in {\em Proceedings of ASIACRYPT 2007}, \ pp.\ 410--426
  (Spinger, Berlin, 2007).

\bibitem{BJT10}
A. Broadbent, S. Jeffrey, and A. Tapp, ``Exact, Efficient and
  Information-Theoretically Secure Voting with an Arbitrary Number of
  Cheaters,'' arXiv:1011.5242 [cs.CR]  (2010).

\bibitem{Boykin:2002PhD}
P. Boykin, {\em Information security and quantum mechanics: security of quantum
  protocols}, Ph.D. thesis, University of California, Los Angeles, 2002.

\bibitem{Christandl:2005cc}
M. Christandl and S. Wehner, ``{Quantum Anonymous Transmissions},''  in {\em
  Proceedings of ASIACRYPT 2005, LNCS 3788} (Springer Berlin Heidelberg,
  Berlin, Heidelberg, 2005), \ pp.\ 217--235.

\bibitem{BBFGT07}
G. Brassard, A. Broadbent, J. Fitzsimons, S. Gambs, and A. Tapp, ``Anonymous
  quantum communication,'' in {\em Proceedings of ASIACRYPT, 2007}, \ pp.\
  460--473  (Springer, Berlin, 2007).

\bibitem{Cai:2013}
X.-Q. Cai and H.-F. Niu, ``Quantum Private Communication with an Anonymous
  Sender,'' Int. J. Theor. Phys. {\bf 52}, 411 (2013).

\bibitem{Nielsen2000}
M.~A. Nielsen and I.~L. Chuang, {\em {Quantum Computation and Quantum
  Information}} (Cambridge, 2000).

\bibitem{Kitaev2003}
A.~Y. Kitaev, ``{Fault-tolerant quantum computation by anyons},'' Annals of
  Physics {\bf 303}, 2 (2003).

\bibitem{Pachos:2012ug}
J.~K. Pachos, {\em {Introduction to Topological Quantum Computation}}
  (Cambridge University Press, 2012).

\bibitem{Barends:2014fk}
R. Barends {\it et~al.}, ``Superconducting quantum circuits at the surface code
  threshold for fault tolerance,'' Nature {\bf 508}, 500 (2014).

\bibitem{QURE}
M. Suchara, J. Kubiatowicz, A. Faruque, F. Chong, C. Lai, and G. Paz-Silva,
  {\em Proceedings of the 31st IEEE International Conference on Computer
  Design} (IEEE, Asheville, North Carolina, 2013), \ p.\ 419.

\bibitem{Demarie:2014jx}
T.~F. Demarie, T. Linjordet, N.~C. Menicucci, and G.~K. Brennen, ``{Detecting
  topological entanglement entropy in a lattice of quantum harmonic
  oscillators},'' New J. Phys. {\bf 16}, 085011 (2014).

\bibitem{JP2012}
J. Pachos, {\em Introduction to $T$opological $Q$uantum $C$omputation}
  (Cambridge University Press, 2012).

\bibitem{Hamma:2005et}
A. Hamma, R. Ionicioiu, and P. Zanardi, ``{Bipartite entanglement and entropic
  boundary law in lattice spin systems},'' Phys. Rev. A {\bf 71}, 022315
  (2005).

\bibitem{Bullock:2007ih}
S.~S. Bullock and G.~K. Brennen, ``{Qudit surface codes and gauge theory with
  finite cyclic groups},'' J. Phys. A: Math. Theor. {\bf 40}, 3481 (2007).

\bibitem{Fowler:2012gc}
A.~G. Fowler, A.~C. Whiteside, and L.~C.~L. Hollenberg, ``{Towards practical
  classical processing for the surface code: Timing analysis},'' Phys. Rev. A
  {\bf 86}, 042313 (2012).

\bibitem{Anwar:2014vr}
H. Anwar, {\em {Towards Fault-Tolerant Quantum Computation with
  Higher-Dimensional Systems}}, Ph.D. thesis, University College London,
  London, 2014.

\bibitem{Brennen:2003ba}
G. Brennen, D. Song, and C. Williams, ``{Quantum-computer architecture using
  nonlocal interactions},'' Phys. Rev. A {\bf 67}, 050302 (2003).

\bibitem{Zhang2008b}
J. Zhang, C. Xie, K. Peng, and P. van Loock, ``{Anyon statistics with
  continuous variables},'' Phys. Rev. A {\bf 78}, 052121 (2008).

\bibitem{Walls2008}
D.~F. Walls and G.~J. Milburn, {\em {Quantum Optics}}, 2nd ed. (Springer,
  Berlin, 2008).

\bibitem{Shannon:1949fb}
C.~E. Shannon, ``{Communication in the Presence of Noise},'' Proc. IRE {\bf
  37}, 10 (1949).

\bibitem{Shannon:1959ge}
C.~E. Shannon, ``{Probability of Error for Optimal Codes in a Gaussian
  Channel},'' Bell System Technical Journal {\bf 38}, 611 (1959).

\bibitem{Cover:2012ub}
T.~M. Cover and J.~A. Thomas, {\em {Elements of Information Theory}} (John
  Wiley {\&} Sons, 2012).

\bibitem{Jaynes:2003vj}
E.~T. Jaynes, {\em {Probability Theory}}, {\em The Logic of Science} (Cambridge
  University Press, 2003).

\bibitem{Urbanke:1998fm}
R. Urbanke and B. Rimoldi, ``{Lattice codes can achieve capacity on the AWGN
  channel},'' IEEE Trans. Inform. Theory {\bf 44}, 273 (1998).

\bibitem{Berens2013}
S. Berens, Master's thesis, Leiden University, 2013.

\bibitem{Genz2009}
A. Genz and F. Bretz, ``Computation of Multivariate Normal and $t$
  Probabilities,''  in {\em Lecture Notes in Statistics} (Springer-Verlag,
  Berlin, Heidelberg, 2009), Vol.~195.

\bibitem{Diehl:2008ha}
S. Diehl, A. Micheli, A. Kantian, B. Kraus, H.~P. B{\"u}chler, and P. Zoller,
  ``{Quantum states and phases in driven open quantum systems with cold
  atoms},'' Nature Phys. {\bf 4}, 878 (2008).

\bibitem{GarZol04}
C.~W. Gardiner and P. Zoller, {\em Quantum Noise}, 3rd  ed. (Springer, 2004).

\bibitem{Jozs94}
R. Jozsa, ``Fidelity for Mixed Quantum States,'' Journal of Modern Optics {\bf
  41}, 2315 (1994).

\bibitem{fidelity13}
G. Spedalieri, C. Weedbrook, and S. Pirandola, ``A limit formula for the
  quantum fidelity,'' Journal of Physics A: Mathematical and Theoretical {\bf
  46}, 025304 (2013).

\bibitem{Wang:2014im}
P. Wang, M. Chen, N.~C. Menicucci, and O. Pfister, ``{Weaving quantum optical
  frequency combs into continuous-variable hypercubic cluster states},'' Phys.
  Rev. A {\bf 90}, 032325 (2014).

\bibitem{Chen:2014jx}
M. Chen, N.~C. Menicucci, and O. Pfister, ``{Experimental Realization of
  Multipartite Entanglement of 60 Modes of a Quantum Optical Frequency Comb},''
  Phys. Rev. Lett. {\bf 112}, 120505 (2014).

\bibitem{Menicucci2011a}
N.~C. Menicucci, ``{Temporal-mode continuous-variable cluster states using
  linear optics},'' Phys. Rev. A {\bf 83}, 062314 (2011).

\bibitem{Yokoyama:2013jp}
S. Yokoyama {\it et~al.}, ``{Ultra-large-scale continuous-variable cluster
  states multiplexed in the time domain},'' Nature Photonics {\bf 7}, 982
  (2013).

\bibitem{Eberle:2010jh}
T. Eberle {\it et~al.}, ``{Quantum Enhancement of the Zero-Area Sagnac
  Interferometer Topology for Gravitational Wave Detection},'' Phys. Rev. Lett.
  {\bf 104}, 251102 (2010).

\bibitem{Mehmet:2011bu}
M. Mehmet, S. Ast, T. Eberle, S. Steinlechner, H. Vahlbruch, and R. Schnabel,
  ``{Squeezed light at 1550 nm with a quantum noise reduction of 12.3 dB},''
  Opt. Express {\bf 19}, 25763 (2011).

\bibitem{PazSilva:2009fa}
G. Paz-Silva, S. Rebi{\'c}, J. Twamley, and T. Duty, ``{Perfect Mirror
  Transport Protocol with Higher Dimensional Quantum Chains},'' Phys. Rev.
  Lett. {\bf 102}, 020503 (2009).

\bibitem{Yurke:1988ih}
B. Yurke {\it et~al.}, ``{Observation of 4.2-K equilibrium-noise squeezing via
  a Josephson-parametric amplifier},'' Physical Review Letters {\bf 60}, 764
  (1988).

\bibitem{CastellanosBeltran:2008cg}
M.~A. Castellanos-Beltran, K.~D. Irwin, G.~C. Hilton, L.~R. Vale, and K.~W.
  Lehnert, ``{Amplification and squeezing of quantum noise with a tunable
  Josephson metamaterial},'' Nature Phys. {\bf 4}, 929 (2008).

\bibitem{Flurin:2012hq}
E. Flurin, N. Roch, F. Mallet, M.~H. Devoret, and B. Huard, ``{Generating
  Entangled Microwave Radiation Over Two Transmission Lines},'' Phys. Rev.
  Lett. {\bf 109}, 183901 (2012).

\bibitem{Bergeal:2012hj}
N. Bergeal, F. Schackert, L. Frunzio, and M.~H. Devoret, ``{Two-Mode
  Correlation of Microwave Quantum Noise Generated by Parametric
  Down-Conversion},'' Phys. Rev. Lett. {\bf 108}, 123902 (2012).

\bibitem{Eichler:2011fc}
C. Eichler {\it et~al.}, ``{Observation of Two-Mode Squeezing in the Microwave
  Frequency Domain},'' Phys. Rev. Lett. {\bf 107}, 113601 (2011).

\bibitem{Wilson:2011ir}
C.~M. Wilson {\it et~al.}, ``{Observation of the dynamical Casimir effect in a
  superconducting circuit},'' Nature {\bf 479}, 376 (2011).

\bibitem{Devoret:2007wo}
M.~H. Devoret, S. Girvin, and R. Schoelkopf, ``{Circuit-QED: How strong can the
  coupling between a Josephson junction atom and a transmission line resonator
  be?},'' Annalen der Physik {\bf 16}, 767 (2007).

\bibitem{Ashhab:2010eh}
S. Ashhab and F. Nori, ``{Qubit-oscillator systems in the ultrastrong-coupling
  regime and their potential for preparing nonclassical states},'' Phys. Rev. A
  {\bf 81}, 042311 (2010).

\bibitem{Allman:2014dn}
M.~S. Allman {\it et~al.}, ``{Tunable Resonant and Nonresonant Interactions
  between a Phase Qubit and LC Resonator},'' Phys. Rev. Lett. {\bf 112}, 123601
  (2014).

\bibitem{Moon:2005jr}
K. Moon and S. Girvin, ``{Theory of Microwave Parametric Down-Conversion and
  Squeezing Using Circuit QED},'' Phys. Rev. Lett. {\bf 95}, 140504 (2005).

\bibitem{Zagoskin:2008kp}
A. Zagoskin, E. Il'ichev, M. McCutcheon, J. Young, and F. Nori, ``{Controlled
  Generation of Squeezed States of Microwave Radiation in a Superconducting
  Resonant Circuit},'' Phys. Rev. Lett. {\bf 101}, 253602 (2008).

\bibitem{YiHuo:2008ke}
W. Yi~Huo and G. Lu~Long, ``{Entanglement and squeezing in solid-state
  circuits},'' New J. Phys. {\bf 10}, 013026 (2008).

\bibitem{Li:2011he}
P.-B. Li and F.-L. Li, ``{Engineering squeezed states of microwave radiation
  with circuit quantum electrodynamics},'' Phys. Rev. A {\bf 83}, 035807
  (2011).

\bibitem{Grimsmo:2017aa}
A.~L. Grimsmo and A. Blais, ``Squeezing and quantum state engineering with
  Josephson travelling wave amplifiers,'' npj Quantum Information {\bf 3}, 20
  (2017).

\bibitem{Yoshikawa2016}
J. ichi Yoshikawa {\it et~al.}, ``Invited Article: Generation of
  one-million-mode continuous-variable cluster state by unlimited time-domain
  multiplexing,'' APL Photonics {\bf 1}, 060801 (2016).

\bibitem{Menicucci2011}
N.~C. Menicucci, S.~T. Flammia, and P. van Loock, ``{Graphical calculus for
  Gaussian pure states},'' Phys. Rev. A {\bf 83}, 042335 (2011).

\bibitem{Menicucci2008}
N.~C. Menicucci, S.~T. Flammia, and O. Pfister, ``{One-Way Quantum Computing in
  the Optical Frequency Comb},'' Phys. Rev. Lett. {\bf 101}, 130501 (2008).

\bibitem{Flammia2009}
S.~T. Flammia, N.~C. Menicucci, and O. Pfister, ``{The Optical Frequency Comb
  as a One-Way Quantum Computer},'' J. Phys. B {\bf 42}, 114009 (2009).

\bibitem{Alexander:2014ew}
R.~N. Alexander, S.~C. Armstrong, R. Ukai, and N.~C. Menicucci, ``{Noise
  analysis of single-mode Gaussian operations using continuous-variable cluster
  states},'' Phys. Rev. A {\bf 90}, 062324 (2014).

\bibitem{Vahlbruch2016}
H. Vahlbruch, M. Mehmet, K. Danzmann, and R. Schnabel, ``Detection of 15 dB
  Squeezed States of Light and their Application for the Absolute Calibration
  of Photoelectric Quantum Efficiency,'' Phys. Rev. Lett. {\bf 117}, 110801
  (2016).

\bibitem{Jiang:2012ja}
L. Jiang, G. He, D. Nie, J. Xiong, and G. Zeng, ``{Quantum anonymous voting for
  continuous variables},'' Phys. Rev. A {\bf 85}, 042309 (2012).

\bibitem{Stace:2009eb}
T. Stace, S. Barrett, and A. Doherty, ``{Thresholds for Topological Codes in
  the Presence of Loss},'' Phys. Rev. Lett. {\bf 102}, 200501 (2009).

\bibitem{Grimmett:1999ur}
G. Grimmett, {\em {Percolation}} (Springer Science {\&} Business Media, 1999).

\bibitem{Borowska:2014ub}
J. Borowska and L. {\L}aci{\'n}ska, ``{Recurrence form for determinant of a
  heptadiagonal symmetric Toeplitz matrix},'' Journal of Applied Mathematics
  and Computational Mechanics {\bf 13}, 19 (2014).

\bibitem{Cinkir:2014jx}
Z. Cinkir, ``{A fast elementary algorithm for computing the determinant of
  Toeplitz matrices},'' Journal of Computational and Applied Mathematics {\bf
  255}, 353 (2014).

\bibitem{Hu:1996vj}
G.~Y. Hu and R.~F. O'Connell, ``{Analytical inversion of symmetric tridiagonal
  matrices},'' J. Phys. A: Math. Gen. {\bf 29}, 1511 (1996).

\bibitem{Elouafi:2014wj}
M. Elouafi, ``{On a relationship between Chebyshev polynomials and Toeplitz
  determinants},'' Applied Mathematics and Computation {\bf 229}, 27 (2014).

\bibitem{AlvarezNodarse:2012fa}
R. {\'A}lvarez-Nodarse, J. Petronilho, and N.~R. Quintero, ``{Spectral
  properties of certain tridiagonal matrices},'' Linear Algebra and its
  Applications {\bf 436}, 682 (2012).

\bibitem{DaFonseca:2005fi}
C.~M. Da~Fonseca and J. Petronilho, ``{Explicit inverse of a tridiagonal k
  Toeplitz matrix},'' Numerische Mathematik {\bf 100}, 457 (2005).

\end{thebibliography}

\vspace{2em}
\hrule
\begin{appendix}
	\section{ Players' covariance matrix (before broadcast) } \label{sec:covmat}

In this section we calculate the covariance matrix %
and associated statistics of the players' measurements of the string momentum operator $\op M$ corresponding to the initial state, that is, \emph{before} any broadcast is sent.

		\subsection{Preparation by measurement of a CV cluster state} \label{sec:prepbymeasurement}

A CV toric-code state can be prepared from a CV cluster state using local measurements, as described in Ref. \cite{Demarie:2014jx}. 		
Given the exact nullifiers for a finitely squeezed CV cluster state on a square lattice (specifically, a weight-1, canonical CV cluster state)~\cite{Menicucci2011}, 
\begin{align}
	\op{\eta}^{\rm CS}_j = \frac{1}{\sqrt{2}} \bigg[ s^{-1} \op q_j + i s \bigg( \op p_j - \sum_{k \in \mathcal N(j)} \op q_k \bigg) \bigg],
\end{align} 
the measured modes lie on the vertices and the face centers of graph for the CV surface-code state, while the unmeasured nullifiers lie on the edges [see Fig.~1(b) and Fig.~\ref{sup:fig:toric}]. Consider an alternating sum of cluster-state nullifiers ~$\op \eta^{\rm CS}_j$ \blk centered on the nodes of a loop  $\primpath$ [e.g.,~every other node left to right through the middle of Fig.~\ref{sup:fig:toric}(a): $\dotsc,72,98,\dotsc$]%
. This sum is also a nullifier of the original CV cluster state. The overlapping $\op q$ terms have canceled, and the sum can be written (up to normalization) as  
\begin{align} \label{eq:CSnulladd} 
\frac{-i}{\sqrt{\abss{\primpath}}} \sum_{e_k\in\primpath}(-1)^k\op{\eta}^{\rm CS}_{k} =\op f-\frac{s}{\sqrt{2\abss{\primpath}}}\sum_{e_k\in\primpath}(\op q_{v^L_k}+\op q_{v^R_k}),
\end{align} 
where $\op q_{v^{L(R)}_k}$ are the position operators for the modes to the left (right) of the edge $e_k$  with respect to~$\primpath$,  and they are located at the faces %
of the CV toric code (e.g.,~nodes $71,73,97,99$). 
We have defined the  string operator~$\op f$ around the loop~$\primpath$,
\begin{align} \label{eq:stringmode}
	 \op f \coloneqq \frac{1}
	 {\sqrt{2\abss \primpath}}
	  \sum_{e\in \primpath}
	  o(e) \left(s\op p_e-is^{-1}\op q_e \right), 
\end{align}
where $\abss{\primpath}$ is the loop length and $o(e)=\pm 1$ if edge $e$ is oriented in the same (opposite) direction as $\primpath$.  

Since these modes are measured in the $\op q$ basis we have a record of their values $\{q_{v^L_k},q_{v^R_k}\}$.  Call the accumulated value 
\begin{align} 
Q=\frac{s}{\sqrt{2\abss{\primpath}}}\sum_{e_k\in\primpath}(q_{v^L_k}+q_{v^R_k}).
\label{QShift} 
\end{align} \blk
Then, the prepared state  satisfies
\begin{align}
\label{eq:stringvacuumcondition}
	(\op f-Q)\ket{\GSs}
	=0
	.
\end{align}
Henceforth, we take
$Q=0$
because the displacement can be accounted for in the protocol by subtracting the value $Q$ when inferring the broadcast message.

\blk

		\subsection{Logical modes of the finitely squeezed CV toric code} \label{sec:logicalmodes}

While string operators (complete loops) are exact logical operators in the case of the qubit~\cite{Kitaev2003}, qudit~\cite{Bullock:2007ih}, and ideal (infinitely squeezed)~\cite{Zhang2008b} CV toric codes, they are only approximately so in the case of a finitely squeezed CV toric code. This is because, as noted in Eqs.~\eqref{eq:nullnoncommute}, finitely squeezed toric-code nullifiers fail to commute with their daggered neighbours. We can, however, identify a set of modes that commute with all toric-code nullifiers and their daggers.

One possible definition of these two logical modes is %
\begin{subequations}
\begin{align}
	\op a_{\logic,\nearrow}
&
\coloneqq
	\frac {1} {\sqrt N} \sum_{e \in \mathcal E} o_\nearrow(e) \op a_e
	,
\\
	\op a_{\logic,\nwarrow}
&
\coloneqq
	\frac {1} {\sqrt N} \sum_{e \in \mathcal E} o_\nwarrow(e) \op a_e
	,
\end{align}
\end{subequations}
where $\mathcal E$ is the set of edges in Fig.~\ref{fig:TCandprotocol}(c), $N$ is the total number of physical modes (note that $\abss{\mathcal E} = N$), and (recalling that the CV toric code is defined on an oriented lattice),
\begin{subequations}
\begin{align}
	o_\nearrow(e)
&
\coloneqq
	\begin{cases}
		+1	& \text{if edge~$e$ is oriented $\uparrow$ or $\rightarrow$},
	\\
		-1	& \text{if edge~$e$ is oriented $\downarrow$ or $\leftarrow$},
	\end{cases}
\\
	o_\nwarrow(e)
&
\coloneqq
	\begin{cases}
		+1	& \text{if edge~$e$ is oriented $\uparrow$ or $\leftarrow$},
	\\
		-1	& \text{if edge~$e$ is oriented $\downarrow$ or $\rightarrow$}.
	\end{cases}
\end{align}
\end{subequations}
The subscript~$\nearrow$ or~$\nwarrow$ on~$o$ is chosen to make this definition intuitive. This results, as can be seen from the orientations of the edges in Fig.~\ref{fig:TCandprotocol}(c), in two operators formed as linear combinations of the physical modes, with signs that alternate along one of the two diagonals and are constant on the other. In fact, the mode shape corresponds to the highest-spatial-frequency standing-wave modes commensurate with the lattice in the two diagonal directions. The two logical mode operators are canonical---i.e., $[\op a_{\logic,i}, \op a^\dagger_{\logic,j}] = \delta_{ij}$, where $i,j \in \{\nearrow, \nwarrow\}$. They satisfy
\begin{align}
	\op a_{\logic,\nearrow} \ket{\GSvac}
=
	\op a_{\logic,\nwarrow} \ket{\GSvac}
=
	0.
\end{align}

By taking linear combinations, we can define operators that have support only on vertical and horizontal edges---i.e.,
\begin{subequations}
\label{eq:frect}
\begin{align}
	\op a_{\logic, \uparrow}
&
\coloneqq
	\frac {1} {\sqrt 2}
	(\op a_{\logic,\nearrow} + \op a_{\logic,\nwarrow})
=
	\sqrt{\frac 2 N} \sum_{e \in \mathcal E_\updownarrow} o_\uparrow(e) \op a_e
	,
\\
	\op a_{\logic, \rightarrow}
&
\coloneqq
	\frac {1} {\sqrt 2}
	(\op a_{\logic,\nearrow} - \op a_{\logic,\nwarrow})
=
	\sqrt{\frac 2 N} \sum_{e \in \mathcal E_\leftrightarrow} o_\rightarrow(e) \op a_e
	,
\end{align}
\end{subequations}
respectively, where the subscript~$\updownarrow$ ($\leftrightarrow$) on~$\mathcal E$ restricts the set to only those edges that are vertical (horizontal), and where the $o$~functions are~$\pm 1$ if the orientation of~$e$ is the same (opposite) of the arrow in the subscript. Examining Fig.~\ref{fig:TCandprotocol}(c), we see that both of these modes have signs alternating in a checkerboard pattern.

The important difference between this situation and that of the qubit~\cite{Kitaev2003}, qudit~\cite{Bullock:2007ih}, or ideal CV~\cite{Zhang2008b} toric code is that the exact logical modes defined in Eqs.~\eqref{eq:frect} are linear combinations of \emph{all} string modes along the same direction. Individual string modes are now \emph{approximate} logical modes, with the approximation improving as the squeezing factor~$s$ increases.

A full description of the finitely squeezed CV toric code will presented in a separate publication. We conclude this subsection by justifying the description of the CV toric-code ground states presented in Sec.~\ref{subsec:finitelysqueezed}.

First, one can explicitly verify that any of the modes defined above commute with all nullifiers and with their daggers---both in the symmetric case [Eqs.~\eqref{eq:nullifiers}] and in the asymmetric case~\cite{Demarie:2014jx}, which is further discussed below. The logical modes are related to the physical modes by a passive transformation, which means that the simultaneous vacuum state of all physical modes is also vacuum in the logical subspace, thereby justifying Eqs.~\eqref{eq:logicalvac} and~\eqref{eq:GSvac}.

We now repeat the analysis of Appendix~\ref{sec:prepbymeasurement}---which applies to the CV toric-code state obtained by measuring a CV cluster state \cite{Demarie:2014jx}---using these logical modes instead of individual string modes.
We find%
\begin{subequations}
\label{eq:flogical}
\begin{align}
	\op f_\nearrow
&
\coloneqq
	\frac {-i} {\sqrt N} \sum_{e_k \in \mathcal E} o_\nearrow(e_k) \op{\eta}^{\rm CS}_{k}
=
	 \frac {1} {\sqrt {2N}} 
	\sum_{e\in \mathcal E}
	o_\nearrow(e) \left(s\op p_e-is^{-1}\op q_e \right)
	,
\\
	\op f_\nwarrow
&
\coloneqq
	\frac {-i} {\sqrt N} \sum_{e_k \in \mathcal E} o_\nwarrow(e_k) \op{\eta}^{\rm CS}_{k}
=
	\frac {1} {\sqrt {2N}} 
	\sum_{e\in \mathcal E}
	o_\nwarrow(e) \left(s\op p_e-is^{-1}\op q_e \right)
	,
\end{align}
\end{subequations}
and therefore
\begin{align}
	\op f_\nearrow \ket{\GSs}
=
	\op f_\nwarrow \ket{\GSs}
=
	0.
\end{align}
Note that by including all unmeasured modes, we have eliminated the dependence on the measurement outcomes [compare with Eq.~\eqref{eq:stringvacuumcondition}]. Also notice that~$\op f_\nearrow$ and~$\op f_\nwarrow$ are merely (up to a phase) squeezed versions of~$\op a_{\logic,\nearrow}$ and~$\op a_{\logic,\nwarrow}$, respectively, with squeezing factor~$s$. This justifies Eq.~\eqref{eq:GSs}.

Finally, note that
\begin{align}
	(\alpha \op a_{\logic,\nearrow} + \beta \op a_{\logic,\nwarrow}) \ket{\GSvac} &= 0,
\\
	(\alpha \op f_\nearrow + \beta \op f_\nwarrow) \ket{\GSs} &= 0,
\end{align}
$\forall \alpha, \beta \in \complex$. Therefore, expressing~$\ket\GSvac$ or~$\ket\GSs$ in a different set of modes within the logical subspace will also have the same form as long as those modes are related to the original ones by a passive transformation.

\blk

		\subsection{General formulation}\label{genform}

We prepare a finitely squeezed CV toric code via measurements on a canonical CV cluster state, as described in Ref. \cite{Demarie:2014jx}. In this case, the CV toric-code face nullifiers are unchanged, but the vertex nullifiers deviate slightly from the symmetric nullifiers defined in Eqs.~\eqref{eq:nullifiers}. These \emph{asymmetric} nullifiers are
\begin{subequations}
\begin{align}
	\hat{\tilde{\eta}}_v & \coloneqq \frac{1}{\sqrt{8}} \bigg[ \sum_{e \in +_v} \left(\tilde{s}\op q_{e}+i\tilde{s}^{-1} \op p_{e} \right) +s^2\tilde{s}^{-1}\sum_{e \in \Diamond_v}   \op q_e \bigg] \, , \\
	\hat{\tilde{\eta}}_f &\coloneqq \frac{1}{\sqrt{8}}\sum_{e \in\square_f} o(e,f) \left(s\op p_{e}-is^{-1}\op q_{e} \right)
\end{align}
\end{subequations}
(with some simple modifications if on a surface with boundary), where $\tilde{s}=\sqrt{5s^2+s^{-2}}$, and $\Diamond_v$ means the diamond shaped loop of next nearest neighbours to the vertex $v$ \cite{Demarie:2014jx}.  
\begin{figure*}[t]
\begin{centering}
\raisebox{1em-\height}{\includegraphics[width=.78\columnwidth]{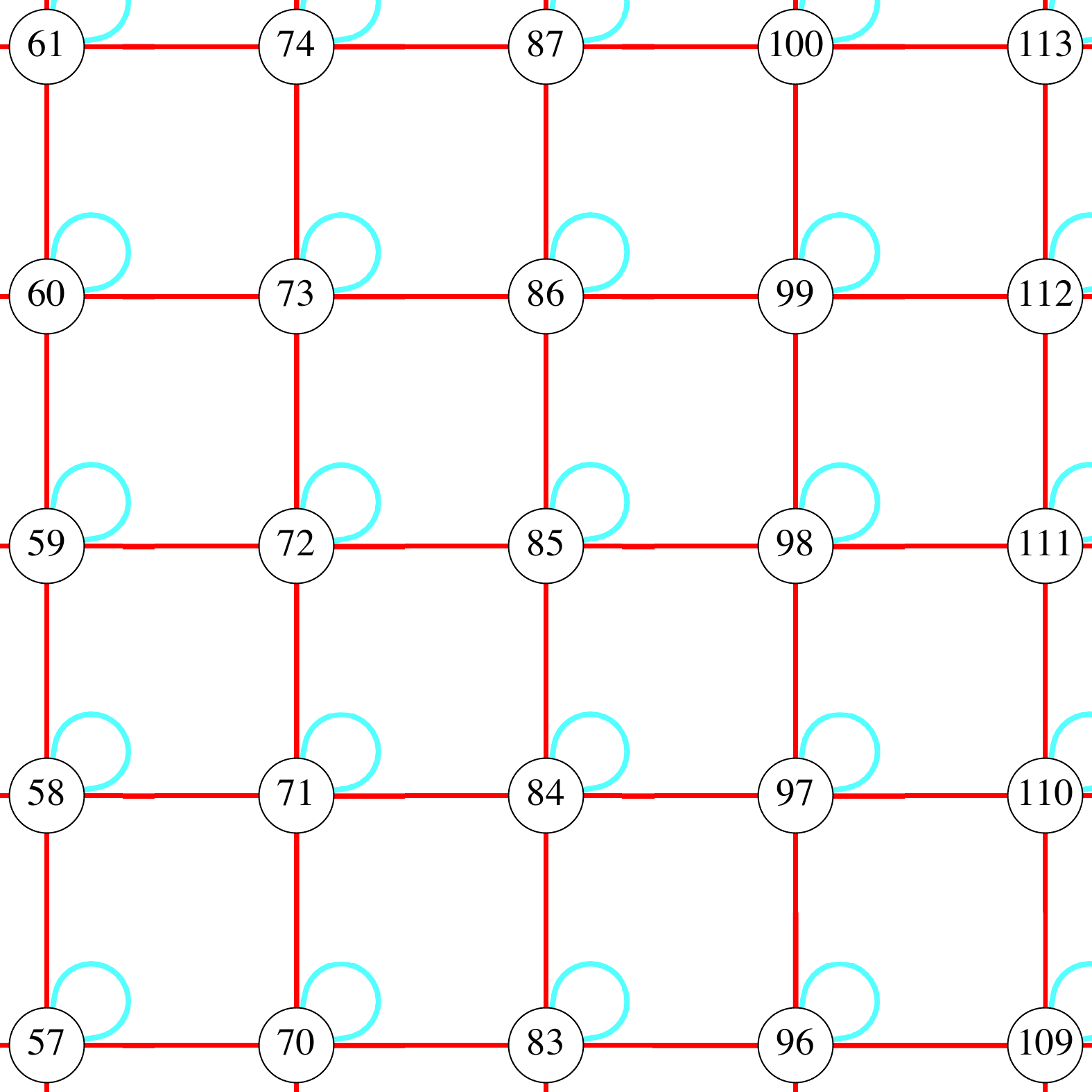}} \hspace{1cm}
\raisebox{1em-\height}{\includegraphics[width=.78\columnwidth]{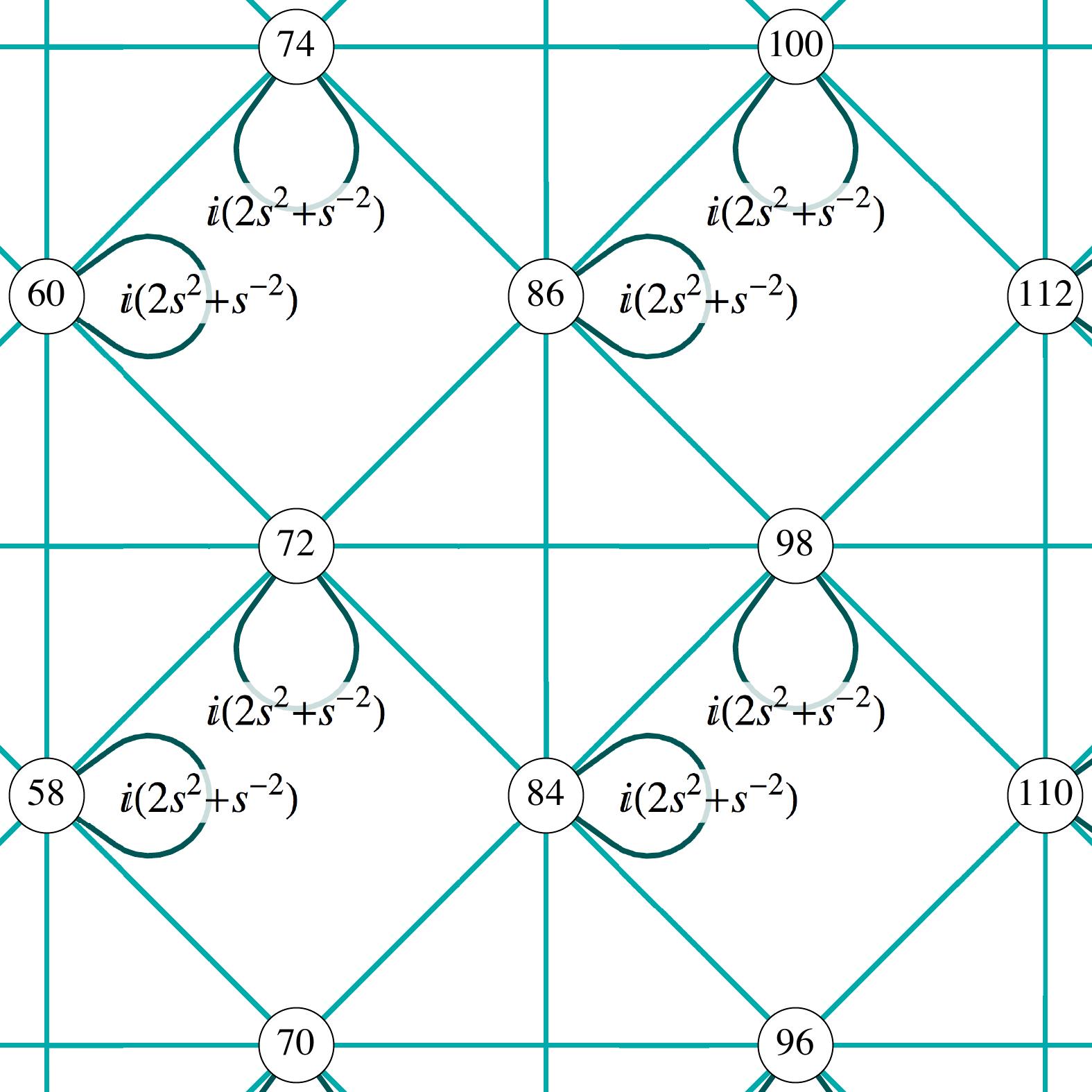}}\\\ \\
~\hspace{.25\columnwidth}{\scriptsize (a)}\hspace{.88\columnwidth}{\scriptsize (b)}
\end{centering}
\caption{\label{sup:fig:toric}Toroidal CV cluster state and toroidal CV surface-code state~$\ket{\GSs}$~\cite{Demarie:2014jx}. \textbf{(a)}~Portion of a CV cluster state with toroidal boundary conditions. Red edges have weight~1, and cyan self-loops have weight~$i s^{-2}$~\cite{Menicucci2011}. \textbf{(b)}~Portion of a CV surface-code state with toroidal boundary conditions (CV toric-code state). Unlabeled edges all have weight $i s^2$. This state is generated by measuring $\op p$ and $\op q$ on the odd nodes of~(a) in a diagonally alternating pattern. The $\op p$ measurements delete the node and produce a criss-cross pattern in~(b) where the node used to be. The $\op q$ measurements just delete the node. (In this case, $\op q$ was measured on nodes $71, 73, 97, 99$;  $\op p$ was measured on the other visible odd-numbered nodes; and so on.)}
\end{figure*}

Figure~6 of Ref.~\cite{Demarie:2014jx} shows the Gaussian graph~\cite{Menicucci2011} for the CV surface code state~$\ket{\GSs}$ created from a canonical CV cluster state, which is also reproduced here in Figure~\ref{sup:fig:toric}(b).
Since its graph~$\mat Z = i\mat U$ is purely imaginary, it directly encodes the $p p$ correlations~\cite{Menicucci2011}: $\avgg{\opvec p \opvec p^\tp} = \frac 1 2 \mat U$.  

When using this state for anonymous broadcasting, $\primpath$ is left to right along one of these horizontal lines---e.g., $\dotsc, 72, 98,\dotsc$ in Figure~\ref{sup:fig:toric}(b). We can write each player's measurement operator~$\op M_j$ along a portion $\primpath_j$ of this path as the inner product between the vector of momentum operators~$\opvec p$ and a normalized indicator vector~$\vec \ell_j = \abss{\primpath_j}^{-1/2} \vec \lambda_j$, where all entries of $\vec \lambda_j$ are~$\pm 1$ or~0. Assuming the width of each wedge is~$w$, then the portion of the string momentum, Eq.~(\ref{Eq:MomWedge}), can be expressed as
\begin{align} \label{eq:MomWedgeVec}
	\op M_j = \vec \ell_j^\tp \opvec p = \frac {1} {\sqrt w} \vec \lambda_j^\tp \opvec p\,.
\end{align}
With respect to the initial state (i.e.,~before any displacements intended to broadcast a message),
\begin{align}
\label{eq:MjMklambda}
	\avgg{\op M_j \op M_k} &= \vec \ell_j^\tp \avgg{\opvec p \opvec p^\tp} \vec \ell_k \nonumber \\
	&= \frac 1 w \vec \lambda_j^\tp \left(\frac 1 2 \mat U\right) \vec \lambda_k \nonumber \\
	&= \frac {1} {2w} \tr (\mat U \vec \lambda_k \vec \lambda_j^\tp)\,.
\end{align}
We can also consider the total string momentum measurement~$\op M = \vec \ell^\tp \opvec p = \abss{\primpath}^{-1/2} \vec \lambda^\tp \opvec p$. Assuming $n$ players and a width-$w$ wedge given to each player,
\begin{align}
	(\Delta M)^2 \coloneqq \avgg{\op M^2} = \frac {1} {2nw} \tr (\mat U \vec \lambda \vec \lambda^\tp)\,.
\end{align}
To illustrate the use of these formulas, it will be instructive to first analyze a simple case.

		\subsection{Simple case: 4-mode CV GHZ state}\label{subsec:GHZ}
\begin{figure}[b]
\begin{centering}
{\scriptsize (a)}\raisebox{1em-\height}{\includegraphics[width=\columnwidth-1em]{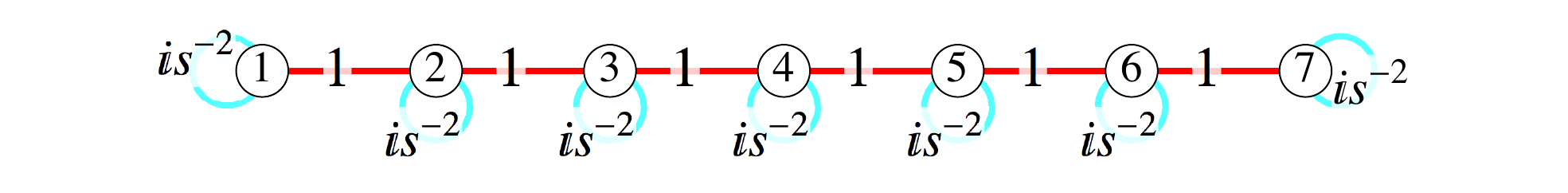}}\\
{\scriptsize (b)}\raisebox{1em-\height}{\includegraphics[width=\columnwidth-1em]{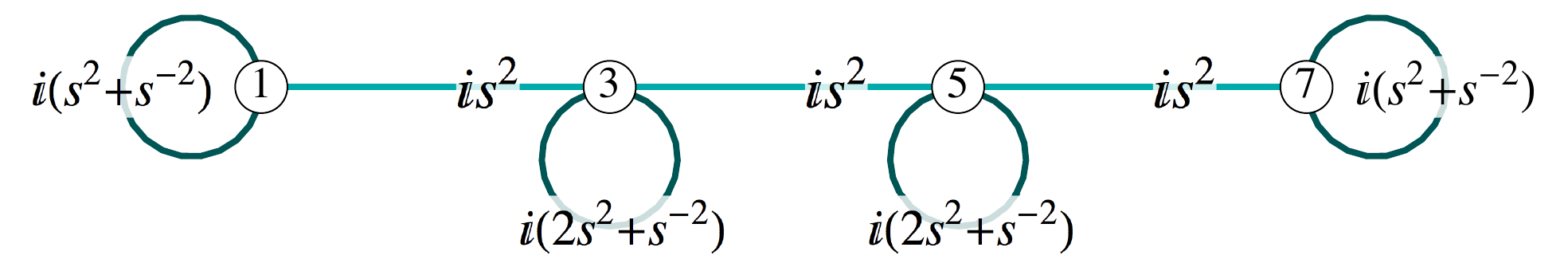}}
\end{centering}
\caption{\label{sup:fig:linear} Graphs for the (a)~linear CV cluster state and (b)~CV GHZ state, with all edge weights labeled explicitly. Measuring $\op p$ on the even nodes in~(a) produces~(b). Notice that the self-loops at the ends of the GHZ state have a different weight from the ones in the middle.}
\end{figure}

Consider the linear CV cluster state in Figure~\ref{sup:fig:linear}(a). By measuring $\op p$ on all even nodes, this state becomes the CV GHZ state whose Gaussian graph~$\mat Z$~\cite{Menicucci2011} is shown in Figure~\ref{sup:fig:linear}(b). Forming its adjacency matrix---also called~$\mat Z$ without ambiguity by taking the nodes in numerical order---we get $\mat Z = i\mat U$ with
\begin{align}
\label{eq:UGHZ}
	\mat U =
	\begin{pmatrix}
		s^2+s^{-2}	& s^2		& 0			& 0		\\
		s^2		& 2s^2+s^{-2}	& s^2		& 0		\\
		0		& s^2		& 2s^2+s^{-2}	& s^2	\\
		0		& 0			& s^2		& s^2+s^{-2}	
	\end{pmatrix}
	\,.
\end{align}
We postulate two players using this state for broadcasting with portions of the string momentum, Eq.~(\ref{eq:MomWedgeVec}), given by
\begin{align}
	\op M_1 &\coloneqq \frac {1} {\sqrt 2}(\op p_1 - \op p_3)\,, \\
	\op M_2 &\coloneqq \frac {1} {\sqrt 2}(\op p_5 - \op p_7)\,.
\end{align}
and total string momentum $\op M = \frac{1}{\sqrt{2}} \big( \op M_1 + \op M_2)$. Therefore,
\begin{align}
	\vec \lambda_1 = 
	\begin{pmatrix}
		1	\\
		-1	\\
		0	\\
		0
	\end{pmatrix}
	\,,\qquad
	\vec \lambda_2 = 
	\begin{pmatrix}
		0	\\
		0	\\
		1	\\
		-1
	\end{pmatrix}
	\,.
\end{align}
The trace in Eq.~\eqref{eq:MjMklambda} is  the Hilbert-Schmidt inner product (entry-wise inner product) between $\mat U$ and $\vec \lambda_j \vec \lambda_k^\tp$. The relevant matrices are
\begin{align}
	\vec \lambda_1 \vec \lambda_1^\tp &= 
	\begin{pmatrix}
		1		& -1		& 0			& 0		\\
		-1		& 1		& 0			& 0		\\
		0		& 0		& 0			& 0		\\
		0		& 0		& 0			& 0		
	\end{pmatrix}
	\,, \\
	\vec \lambda_2 \vec \lambda_2^\tp &= 
	\begin{pmatrix}
		0		& 0		& 0			& 0		\\
		0		& 0		& 0			& 0		\\
		0		& 0		& 1			& -1		\\
		0		& 0		& -1			& 1		
	\end{pmatrix}
	\,, \\
	\vec \lambda_1 \vec \lambda_2^\tp &= 
	\begin{pmatrix}
		0		& 0		& 1			& -1		\\
		0		& 0		& -1			& 1		\\
		0		& 0		& 0			& 0		\\
		0		& 0		& 0			& 0		
	\end{pmatrix}
	\,.
\end{align}
Taking entry-wise inner products of these with $\mat U$, we find the wedge-wise variances,
\begin{align}
	\avgg{\op M_1^2} = \avgg{\op M_2^2} &= \frac 1 4 \bigl[(s^2+s^{-2}) + (2s^2+s^{-2}) - 2s^2 \bigr] \nonumber \\
	&= \frac {s^2} {4} + \frac {1} {2s^2}\,,
\end{align}
and inter-wedge covariances,
\begin{align} \label{eq:WedgeCov}
	\avgg{\op M_1 \op M_2} = \avgg{\op M_2 \op M_1} &= \frac {-s^2} {4}\,.
\end{align}

The total measurement~$\op M$ has $\vec \lambda = \vec \lambda_1 + \vec \lambda_2$. Therefore,
\begin{align}
\label{eq:lambdalambda}
	\vec \lambda \vec \lambda^\tp =
	\begin{pmatrix}
		1		& -1		& 1			& -1		\\
		-1		& 1		& -1			& 1		\\
		1		& -1		& 1			& -1		\\
		-1		& 1		& -1			& 1		
	\end{pmatrix}
	\,,
\end{align}
and the resultant entry-wise inner product with $\mat U$ is the sum of the diagonal of $\mat U$ minus all entries on the sub- and superdiagonals:
\begin{align}
	(\Delta M)^2 = \avgg{\op M^2} &= \frac {1} {8} \bigl[2(s^2 + s^{-2}) + 2(2s^2 + s^{-2}) - 6s^{2} \bigr] \nonumber \\
	&= \frac {1} {2s^2}\,.
\end{align}
Notice that  the large-variance terms ($\sim s^2$) cancel in this sum due to the covariances between the wedges, Eq.~(\ref{eq:WedgeCov}). (The fact that the self-loops at the ends are different from those in the center of the chain is required for this cancelation to happen.) Therefore, the total string momentum measurement has a small variance even though individual players' measurements have a large variance---this is the essence of the anonymous broadcasting protocol.

		\subsection{CV toric-code state} \label{subsec:toric}

We now return to the case of the toric-code state shown in Figure~\ref{sup:fig:toric}(b). We assume a general scenario of $n$ players, each of whom possesses a slice of the torus of width~$w$. Because of the toroidal boundary conditions, $nw$ must be even, and we assume it is not trivially small (i.e., $nw \geq 4$).

For illustration, we start with the concrete example of $w=4$. Then,
\begin{align}
	\vec \lambda_j =
	\begin{pmatrix}
		0 & \cdots & 0 & 1 & -1 & 1 & -1 & 0 & \cdots & 0
	\end{pmatrix}
	^\tp\,,
\end{align}
where the nodes with nonzero entries are numbered along $\primpath$. Since any node not along $\primpath$ corresponds to a 0 in all of the $\vec \lambda_j$, we can consider just the induced subgraph of $\mat U$ restricted to $\primpath$---in other words, the submatrix of $\mat U$ restricted to the nodes along $\primpath$.

Inspection reveals that along $\primpath$, $\mat U$ for the toric code [Figure~\ref{sup:fig:toric}(b)] is exactly like that of the GHZ state [Figure~\ref{sup:fig:linear}(b)] except at the ends, where there is an extra edge connecting the two endpoints and self-loops of weight $2s^2 + s^{-2}$ instead of $s^2 + s^{-2}$. Continuing with the example above (and omitting zeros),
\begin{align}
\label{eq:lambdablockeven}
	\vec \lambda_j \vec \lambda_k^\tp =
	\begin{pmatrix}
		&&&&&&& \\
		&& 1 & -1 & 1 & -1 & \\
		&& -1 & 1 & -1 & 1 & \\
		&& 1 & -1 & 1 & -1 & \\
		&& -1 & 1 & -1 & 1 & \\
		&&&&&&&
	\end{pmatrix}
	\,,
\end{align}
with the size of the blank padding on each side (representing zeros) left unspecified but determined by $j$ and $k$.

The relevant part of $\mat U$ is \emph{circulant} tridiagonal (nodes numbered according to $\primpath$) with all diagonal entries~$2s^2 + s^{-2}$ (no difference at the ends because of periodicity) and all sub- and superdiagonal entries (continued in a circulant fashion) equal to~$s^2$:
\begin{align}
\label{eq:Ucirculant}
	\mat U \mapsto
	\begin{pmatrix}
		a	& s^2		& 		&			& s^2		\\
		s^2			& a	& s^2	&						\\
					& \ddots			& \ddots	& 	\ddots				\\
					& 			& s^2	& a	& s^2		\\
		s^2			& 			& 		& s^2		& a
	\end{pmatrix}
	\,,%
\end{align}
where $a=2s^2+s^{-2}$, nodes are again ordered according to their appearance along $\primpath$, and $\mapsto$ indicates that only the relevant part of the full $\mat U$ is shown [cf.\ Eq.~\eqref{eq:UGHZ}].

When $j=k$, the ${4\times 4}$ block of $\pm 1$ in Eq.~\eqref{eq:lambdablockeven} is on the diagonal, and thus only the three innermost diagonals of that block matter when taking the entry-wise inner product with~$\mat U$. Therefore, for $w=4$, $\avgg{\op M_j^2} = \frac 1 8 [4(2s^2 + s^{-2}) - 6s^2]$. When $j - k = \pm 1 \pmod n$, then the only entry that matters is the $-1$ in the upper right or bottom left of the block, and thus $\avgg {\op M_j \op M_{j\pm 1}} = \frac 1 8(-s^2)$. Analogous results hold for other even values of~$w$, but we will postpone the general formula until we consider the odd case.

When $w$ is odd, the form of the number block in Eq.~\eqref{eq:lambdablockeven} differs depending on whether $j - k$ is even or odd. This is because adjacent measurement operators have opposite sign configurations when adding up the individual $\op p$ operators. Using $w=3$ as an example,
\begin{align}
\label{eq:lambdablockodd}
	\vec \lambda_j \vec \lambda_{j+\text{even}}^\tp &=
	\begin{pmatrix}
		&&&&&& \\
		&& 1 & -1 & 1 &  \\
		&& -1 & 1 & -1 &  \\
		&& 1 & -1 & 1 &  \\
		&&&&&&
	\end{pmatrix}
	\,, \\
	\vec \lambda_j \vec \lambda_{j+\text{odd}}^\tp &=
	\begin{pmatrix}
		&&&&&& \\
		&& -1 & 1 & -1 &  \\
		&& 1 & -1 & 1 &  \\
		&& -1 & 1 & -1 &  \\
		&&&&&&
	\end{pmatrix}
	\,,
\end{align}
with the size of the blank padding on each side (representing zeros) left unspecified but determined by $j$ and $k$.
Notice that, once again, for the same reasons as for even $w$, only the cases where $j=k$ or $j-k=\pm1 \pmod n$ matter, and now the pattern for both even and odd $w$ is clear (and the same in both cases):
\begin{align} 
	\avgg{\op M_j^2} &= \frac {1} {2w} \bigl[ w(2s^2+s^{-2}) - 2(w-1)s^2 \bigr] \nonumber \\
	&= \frac {1} {2s^2} + \frac {s^2} {w}\,, \label{eq:playervariance} \\
	\avgg{\op M_j \op M_{j\pm 1}} &= \frac {-s^2} {2w}\,, \label{eq:playercovariance}
\end{align}
where the $\pm 1$ is mod~$n$. These are the pre-broadcast covariances of the players' measurement operators using a toric-code state. They also hold for the GHZ state with periodic boundary conditions, which is a special case of the torus.

The total measurement~$\op M$  has a matrix~$\vec \lambda \vec \lambda^\tp$ whose nonzero block is ${nw \times nw}$ and of the same form as Eq.~\eqref{eq:lambdalambda}. Notice that in order to get the periodicity to match up, $nw$ must be even. Examining the form of $\mat U$ in Eq.~\eqref{eq:Ucirculant}, we see that we must add the diagonal of $\mat U$ and subtract its sub- and superdiagonals, including their circulant extensions (the entries in the corners). %
Therefore, we have the general result
\begin{align}
\label{eq:DeltaM}
	(\Delta M)^2 = \avgg{\op M^2} &= \frac {1} {2nw} \bigl[ nw(2s^2+s^{-2}) -2nw(s^2) \bigr] \nonumber \\
	&= \frac {1} {2s^2}\,,
\end{align}
which holds for all $n$ and $w$ (with $nw \ge 4$ and even).

		\subsection{CV surface-code state with open boundaries} \label{subsec:open}

The calculations of sender anonymity and broadcast channel capacity assume a toric-code state, whose results were presented above. The optical implementation (Section~\ref{sec:implementation}), however, proposes implementing the protocol using surface-code states with open boundaries instead. Here we show that this sort of resource also works. 

\begin{figure}[t]
\begin{centering}
{\scriptsize (a)}\raisebox{1em-\height}{\includegraphics[width=\columnwidth-1em]{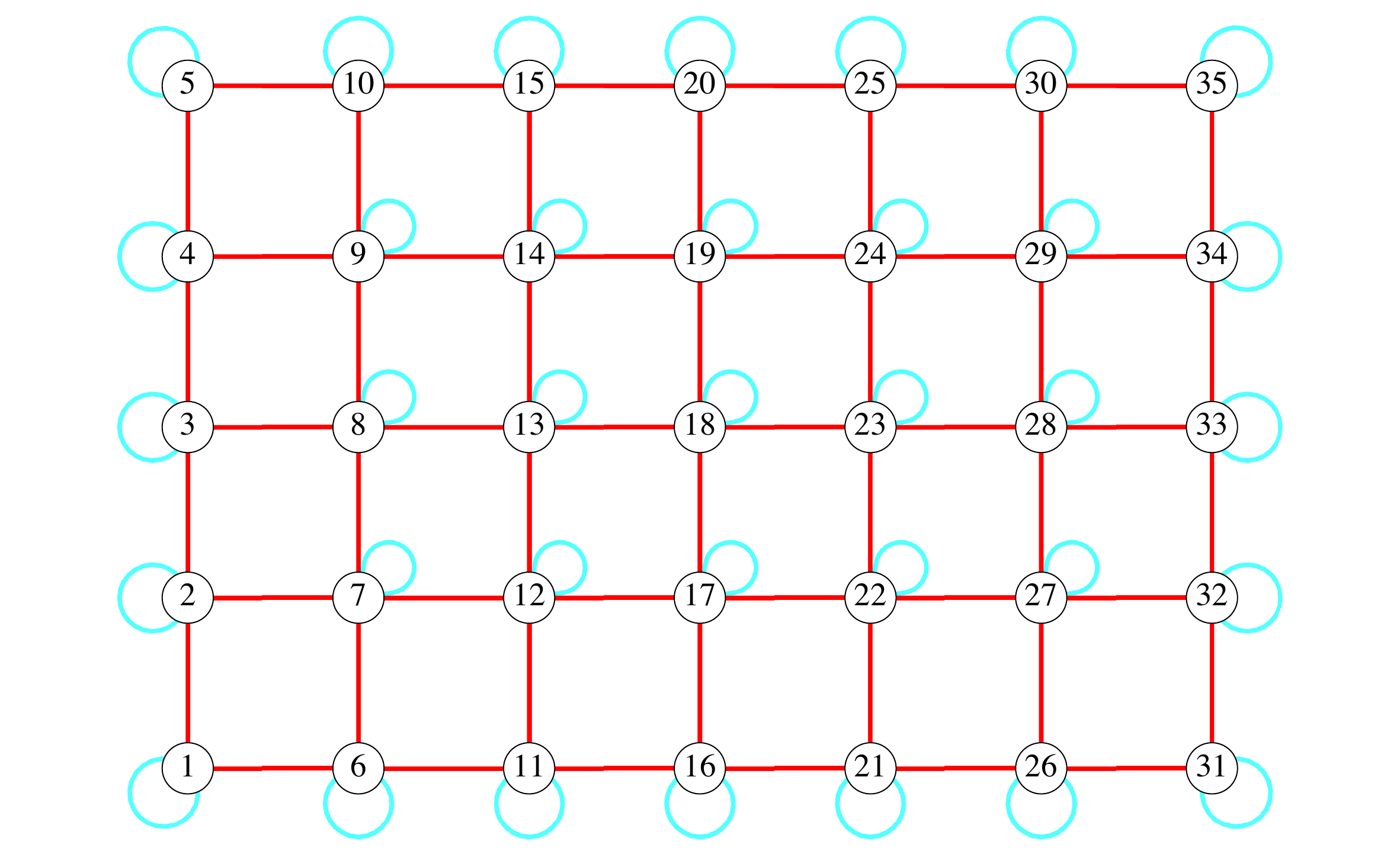}}\\
{\scriptsize (b)}\raisebox{1em-\height}{\includegraphics[width=\columnwidth-1em]{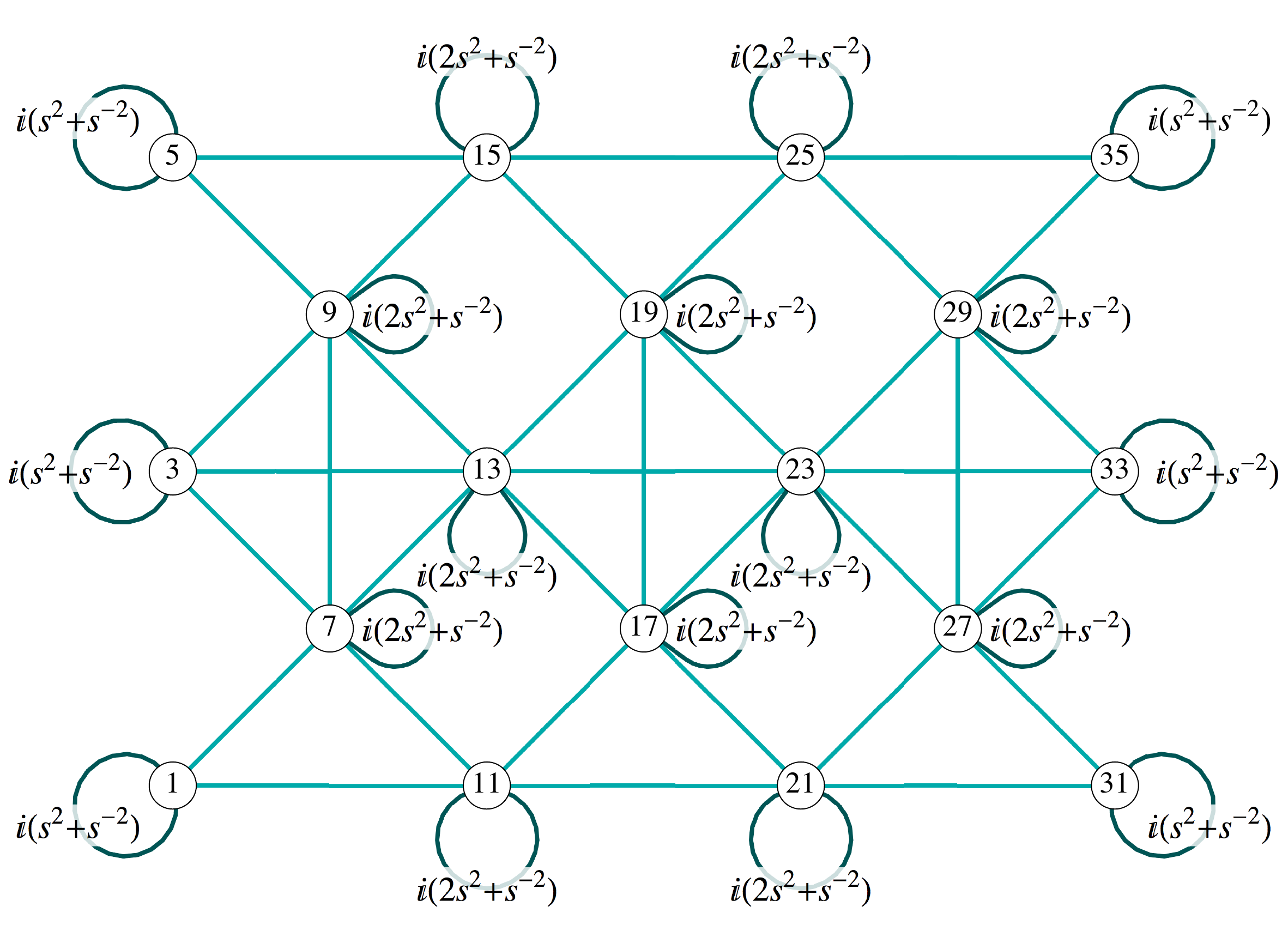}}
\end{centering}
\caption{\label{sup:fig:open}Open-boundary CV cluster state and CV surface-code state. \textbf{(a)}~CV cluster state with open boundaries. Red edges have weight~1, and cyan self-loops have weight~$i s^{-2}$~\cite{Menicucci2011}. \textbf{(b)}~CV surface-code state with smooth boundaries on the top and bottom and with rough boundaries on the left and right. Unlabeled edges all have weight $i s^2$. Starting from~(a), the smooth boundaries are generated by measuring $\op p$ on nodes $6, 16, 26, 10, 20, 30$. The rough boundaries are generated by measuring $\op q$ on nodes $2, 4, 32, 34$. An alternating pattern of $\op p$ and $\op q$ measurements on all remaining even nodes completes the transition to the surface-code state. (The terms `smooth' and `rough' are chosen by convention to visually match the boundaries of the resulting graph.) Also notice that the three horizontal lines extending the full width of~(b) have the same weights as the CV GHZ state from Figure~\ref{sup:fig:linear}(b).}
\end{figure}

The open-boundary surface-code state is shown in Figure~\ref{sup:fig:open}(b), where the top and bottom are `smooth' boundaries, and the left and right are `rough' boundaries, with terminology chosen by convention because of their visual representation in the graph. We can choose $\primpath$ to be any of the three horizontal lines of nodes in that graph that stretch all the way from the left boundary (rough) to the right boundary (also rough)---e.g.,~3, 13, 23, 33. Alice will apply her displacements along $\dualpath$, which could be, for instance, 11, 13, 15, or any of the vertical lines parallel to that one and that stretch all the way from the bottom boundary (smooth) to the top boundary (also smooth).

Notice that the self-loops at the rough boundaries [Figure~\ref{sup:fig:open}(b)] are like the endpoints of the CV GHZ state [Figure~\ref{sup:fig:linear}(b)]. In fact, by the same logic as in the toric-code case above, the only part of $\mat U$ that will matter is the submatrix of the full $\mat U$ limited to the nodes along $\primpath$. This now has the exact same form as the $\mat U$ for the GHZ state, which is given in Eq.~\eqref{eq:UGHZ}. For  $n$ players, the surface-code state is divided into $n$ vertical slices, each with arbitrary width $w$ (with $nw \geq 4$ and even). Then, the matrix $\mat{U}$ becomes
\begin{align}
\label{eq:Uopen}
	\mat U \mapsto
	\begin{pmatrix}
		b	& s^2		& 		&			& 		\\
		s^2			& a	& s^2	&						\\
					& \ddots			& \ddots	& 	\ddots				\\
					& 			& s^2	& a	& s^2		\\
					& 			& 		& s^2		& b
	\end{pmatrix}
	\,,%
\end{align}
where $a=2s^2+s^{-2}$ and $b=s^2+s^{-2}$, and $\mapsto$ again indicates that only the relevant part of $\mat U$ is displayed. Notice the two differences between this and Eq.~\eqref{eq:Ucirculant}: In Eq.~\eqref{eq:Uopen}, the first and last diagonal entries are different from the rest, and the isolated corner entries are missing.

Using the same arguments as above, we have the following variances within each player's slice and the inter-slice covariances:
\begin{align}
	\avgg{\op M_1^2} = \avgg{\op M_n^2} &= \frac {1} {2w} \bigl[ (w-1)(2s^2+s^{-2}) \nonumber \\
	&\qquad + (s^2+s^{-2})- 2(w-1)s^2 \bigr] \nonumber \\
	&= \frac {1} {2s^2} + \frac {s^2} {2w}\,, \\
	\avgg{\op M_j^2} &= \frac {1} {2w} \bigl[ w(2s^2+s^{-2}) - 2(w-1)s^2 \bigr] \nonumber \\
	&= \frac {1} {2s^2} + \frac {s^2} {w}\,, \\
	\avgg{\op M_j \op M_{j\pm 1}} &= \frac {-s^2} {2w}\,,
\end{align}
where $2\le j \le n-1$. Notice that the $\pm 1$ is no longer mod~$n$. Also,
\begin{align}
	(\Delta M)^2 = \avgg{\op M^2} &= \frac {1} {2nw} \bigl[ (nw-2)(2s^2+s^{-2}) \nonumber \\
	&\qquad + 2(s^2+s^{-2}) - 2(nw-1)s^2 \bigr] \nonumber \\
	&= \frac {1} {2s^2}\,.
\end{align}
In this case, the noise of the broadcast message is the same, $(\Delta M)^2 = \frac {1} {2s^2}$, which means the channel capacity is the same (Section~\ref{sec:capacity}). But now players~1 and $n$ are more at risk of being discovered if one of them is the broadcaster. This is because the local noise in their measurement outcomes is less than that of the other players, and it is this local noise that hides the fact that any individual player has broadcast a message (Section~\ref{sec:anonymity}).

One might be tempted to think that making the end slices (1 and $n$) narrower, with a width of~$\frac w 2$ instead of $w$, could make the local noise the same for all players. This is true---but misleading. The reason for this is that if player 1 or $n$ wanted to broadcast a message~$r$, her measurement outcome would be displaced further than would that of players $2,\dotsc,n-1$ if one of them instead had broadcast the same message---in fact, further by a factor of $\sqrt 2$ [see Eq.~\eqref{eq:shift}]. This means that the variance of that displacement is twice what it would be had she used a full $w$-width slice. This effectively nullifies the advantage of increased local noise in the narrower slice. Either way, the local signal-to-noise ratio (which governs the risk of broadcaster discovery) is approximately twice what it would be for any of the other players wishing to broadcast the same message. Thus, there is no advantage to using narrower slices at the ends.

		\section{Wedge width in Figure~5} \label{sec:width}

The results summarized in Fig.~\ref{fig:results} assume that the players have received wedges of width $w=6$. Here we justify this choice.

Assume that in addition to the dissipative error mitigation proposed in the main text, one can also perform measurements of the number of excitations in the nullifiers. A detected excitation indicates an error in the code (a jump out of the code space) in the neighborhood of that nullifier. We then logically tag that location as a part of the code to be avoided---effectively declaring the modes in that neighborhood lost completely. 
This conservative choice allows us to steer clear of detected errors altogether.

For rates of lost (i.e.,~error-tagged) nodes below the toric-code error tolerance rate of 50\% (%
error per mode $p_{\rm err} = \frac 1 2$ per physical operation), as derived from the percolation threshold for a square lattice \cite{Stace:2009eb}, paths can be found that connect the lattice along homologically non-trivial loops. Communication between players restricts the allowable density of errors and defines a lower bound for the width of each wedge. For occupation probability $p$ below the percolation threshold $p_c$, the probability that there is a cluster of radius $r$ in the percolation model is given by $p_{\rm cluster}(r)\approx e^{-r\abs{p-p_c}^{\nu}}$ where $\nu$ is the critical exponent \cite{Grimmett:1999ur}.  For bond percolation on a square lattice in two dimensions, $p_c=\frac 1 2$ and $\nu= \frac 4 3$, so the probability the protocol fails due to these errors is
\begin{align}
	p_{\rm fail}\approx e^{-\frac{w}{2}\abs{p_{\rm err}-\frac 1 2}^{4/3}}\,.
\end{align}
Hence, for a target $p_{\rm fail}$, we have
\begin{align}
	w\geq \frac{2\log(p_{\rm fail}^{-1})}{\abs{p_{\rm err}-\frac 1 2}^{4/3}}\,.
\end{align}

Assume errors can be monitored, for instance using the protocol described in the main text. Then, if one of the players measures a percolated cluster of errors on her wedge, she can announce an abort warning to the others.  The whole protocol can then be retried, and the probability of failure after $k$ attempts is $p_{\rm fail}^k$.  Say we fix $p_{\rm fail}=1/e$, implying\begin{align}
	w \ge \frac {2} {\abs{p_{\rm err}- \frac 1 2}^{4/3}}\,.
\end{align}
Then, assuming an error rate $p_{\rm err}<0.06$, a wedge width of $w=6$ will suffice. This percolation argument also assumes a circumference of the wedge around the same size.

\section{Mathematical results} \label{sec:math}

Here we provide mathematical results that are used in Sections~\ref{sec:capacity} and~\ref{sec:anonymity}.

\subsection{Gaussian distributions: notation and entropic properties}

We adopt the following notation for a random variable~$X$ with instantiations~$x \in \reals$ distributed according to a Gaussian (normal) distribution with mean~$\avg X = \mu$ and variance~$\var(X) = \avg {(X-\mu)^2} = \sigma^2$:
\begin{align}
\label{eq:Gauss}
	X \sim N_{\mu,\sigma^2}(x) = \frac {1} {\sqrt{2\pi\sigma^2}} \exp\left[ -\frac{(x-\mu)^2}{2\sigma^2} \right]\,.
\end{align}
This can easily be extended to a random column vector~$\vec X$ with instantiations~$\vec x \in \reals^n$ distributed according to a multivariate Gaussian with mean~$\avg{\vec X} = \vec \mu$ and covariance matrix~$\cov(\vec X) = \avg{(\vec X - \vec \mu) (\vec X - \vec \mu)^\tp} = \mat \Sigma > 0$:
\begin{align}
\label{eq:Gaussmulti}
	\vec X &\sim N_{\vec \mu,\mat \Sigma}(\vec x) \nonumber \\
	&= \frac {1} {\sqrt{\det (2\pi \mat \Sigma)}} \exp\left[ -\frac 1 2 (\vec x-\vec \mu)^\tp \mat \Sigma^{-1} (\vec x-\vec \mu) \right]\,.
\end{align}
The entropy of the univariate Gaussian is
\begin{align}
\label{eq:Gaussent}
	H(X) &= \avg{-\log N_{\mu,\sigma^2}(X)} %
	=  \frac 1 2 \log(2\pi e \sigma^2)\,.
\end{align}
Note that we leave the base unspecified. Therefore, all entropies in this document are expressed in bits if the log based is 2, in nats if the log base is $e$, etc. Its multivariate generalization is
\begin{align}
\label{eq:Gaussmultient}
	H(\vec X) &= \avg{-\log N_{\vec \mu,\mat \Sigma}(\vec X)} %
	=  \frac 1 2 \log \det(2\pi e \mat \Sigma)\,.
\end{align}
For any random vector~$\vec Y$---not necessarily Gaussian---with mean~$\vec \mu$ and covariance~$\mat \Sigma$, its entropy is bounded from above by the entropy of a Gaussian-distributed random vector with the same covariance. In other words,
\begin{align}
\label{eq:entupper}
	H(\vec Y) \leq \frac 1 2 \log \det(2\pi e \mat \Sigma) = H(\vec X)\,.
\end{align}

\subsection{Special cases of symmetric, tridiagonal, toeplitz/circulant matrices}

Consider the two ${n\times n}$ matrices%
\begin{align}
\label{eq:Tn}
	\mat T_n(x) &\coloneqq
	\begin{pmatrix}
		x & 1 \\
		1 & x & 1 \\
		& 1 & x & 1 \\
		&& \ddots & \ddots & \ddots \\
		&&&1 & x & 1 \\
		&&&& 1 & x & 1 \\
		&&&& & 1 & x \\
	\end{pmatrix}
\intertext{and}
\label{eq:Cn}
	\mat C_n(x) &\coloneqq
	\begin{pmatrix}
		x & 1 &&&&&1\\
		1 & x & 1 \\
		& 1 & x & 1 \\
		&& \ddots & \ddots & \ddots \\
		&&&1 & x & 1 \\
		&&&& 1 & x & 1 \\
		1&&&& & 1 & x \\
	\end{pmatrix}
	\,,
\end{align}
with constant diagonal bands understood and missing entries taken to be~0. The notation is chosen because $\mat T_n(x)$ is a Toeplitz matrix and $\mat C_n(x)$ is its circulant counterpart. These matrices are uniquely defined for ${n\ge 3}$. We can complete the definition for all~$n \in \natnums$ by also defining
\begin{align}
\label{eq:T1C1}
	\mat T_1(x) = \mat C_1(x) &\coloneqq
	\begin{pmatrix}
		x
	\end{pmatrix}
	\,, \\
\label{eq:T2C2}
	\mat T_2(x) = \mat C_2(x) &\coloneqq
	\begin{pmatrix}
		x & 1 \\
		1 & x
	\end{pmatrix}
	\,.
\end{align}
Now let us consider their determinants.

Define $t_n(x) \coloneqq \det \mat T_n(x)$. Using the cofactor expansion of the determinant, we see that the following recurrence relation holds for $n \ge 3$~\cite{Borowska:2014ub,Cinkir:2014jx}:
\begin{align}
	t_n(x) = xt_{n-1}(x) - t_{n-2}(x)\,.
\end{align}
Since $t_2(x) = x^2-1$ and $t_1(x) = x$ by direct calculation, we see that this recurrence relation also holds for $n=2$ if we choose $t_0(x) \coloneqq 1$. These are exactly the recurrence relation and initial conditions for the Chebyshev polynomials of the second kind~$U_n(\tfrac x 2)$. Therefore,
\begin{align}
\label{eq:detTn}
	\det \mat T_n(x) = t_n(x) = U_n \left(\frac x 2 \right)\,.
\end{align}
This result also agrees with the literature~\cite{Hu:1996vj,Elouafi:2014wj,AlvarezNodarse:2012fa,DaFonseca:2005fi} after applying properties of Chebyshev polynomials.

Define $c_n(x) \coloneqq \det \mat C_n(x)$. A cofactor expansion for $n\ge 3$ relates this to the result for the Toeplitz case:
\begin{align}
	c_n(x) &= xt_{n-1}(x) - 2[t_{n-2}(x) + (-1)^n]\,.
\end{align}
Plugging in Eq.~\eqref{eq:detTn} and using properties of Chebyshev polynomials gives
\begin{align}
	c_n(x) %
	&= 2(-1)^n \left[T_n \left(-\frac x 2\right) -1 \right]\,,
\end{align}
where $T_n$ is the $n$th-order Chebyshev polynomial of the first kind, valid for $n\ge 3$. Note that $c_2(x) = t_2(x)$ and $c_1(x) = t_1(x)$. Therefore,
\begin{align}
\label{eq:detCn}
	\det \mat C_n(x) &= c_n(x) \nonumber \\
	&=
	\begin{cases}
		U_n \left(\frac x 2 \right) & \text{if $n \in \{1,2\}$}, \\
		2(-1)^n \left[T_n \left(-\frac x 2\right) -1 \right] & \text{if $n\ge 3$}.
	\end{cases}
\end{align}

Now consider a perturbed version of the circulant matrix above:
\begin{align}
\label{eq:Cshift}
	\mat C_n(x,a) \coloneqq 
	\begin{pmatrix}
		x+a & 1 &&&&&1\\
		1 & x & 1 \\
		& 1 & x & 1 \\
		&& \ddots & \ddots & \ddots \\
		&&&1 & x & 1 \\
		&&&& 1 & x & 1 \\
		1&&&& & 1 & x \\
	\end{pmatrix}
	\,.
\end{align}
Cofactor evaluation of its determinant gives
\begin{align}
	\det \mat C_n(x,a) &= \det \mat C_n(x) + a \det \mat T_{n-1}(x)\,.
\end{align}
Specializing to $n\ge 3$ evaluates this to
\begin{align}
\label{eq:detCshift}
	\det \mat C_n(x,a) &= 2(-1)^n \left[T_n \left(-\frac x 2\right) -1 \right] + a U_{n-1} \left(\frac x 2 \right) \nonumber \\
	&= 2(-1)^n \left(1+\frac a n \frac {\partial} {\partial x} \right) \left[T_n \left(-\frac x 2\right) -1 \right]\,.
\end{align}
Notice that this means
\begin{align}
\label{eq:detCshiftgeneral}
	\det \mat C_n(x,a) &= \left(1+\frac a n \frac {\partial} {\partial x} \right) \det \mat C_n(x)\,,
\end{align}
which can be also be verified using Jacobi's identity. Direct evaluation for $n=1$ and $n=2$ show that Eq.~\eqref{eq:detCshiftgeneral} is also valid for those cases and therefore valid for all $n \in \natnums$. %

\section{Dissipative Gaussian Dynamics} \label{sec:GDynamics}

For $N$ modes undergoing Gaussian dynamics, the master equation describes the evolution of the means and covariance matrix~$\mat{\Sigma}$. Defining a column vector of stacked quadrature operators
\begin{align}
	\opvec{r} \coloneqq
	\begin{pmatrix}
		\opvec{q} \\ \opvec{p}
	\end{pmatrix}
	,
\end{align}
the commutation relations can be expressed as
\begin{align}
	[\opvec{r}, \opvec{r}^\tp ] = \opvec{r} \opvec{r}^\tp - ( \opvec{r} \opvec{r}^\tp ){}^\tp = i \mat{\Omega}
,
\end{align}
where ${}^\tp$ indicates matrix transpose (see Ref. \cite{Menicucci2011} for more details on this notation). The matrix $\mat{\Omega}$ is known as the \emph{symplectic form}:
	\begin{align}
		\mat{\Omega} = 
		\begin{pmatrix}
			\mat{0} &  \mat \id  \\
			-\mat \id & \mat{0}
		\end{pmatrix},
	\end{align}
where $\mat \id$ is the  $N \times N$  identity matrix.
For dissipative evolution given by a Markov master equation, Gaussianity is preserved under two conditions. First, the Hamiltonian must have the quadratic form
	\begin{align}
		\op H = \frac 1 2 \opvec{r}^\tp \mat{G} \opvec{r},
	\end{align}
expressed in terms of the symmetric, real matrix $\mat{G} \in \reals^{2N \times 2N}$. Second, the jump operators that describe coupling to $M$ baths must be linear in the mode operators. The jump operator that couples to bath $k$ therefore must have the form
	\begin{align} \label{Eq::JumpOps}
		\op L_k = & \sum_{j = 1}^N \big( Q_{kj} \op q_j + P_{kj} \op p_j \big).
	\end{align}
We collect all $M$ of these jump operators into the vector of operators denoted
	\begin{align} \label{Eq::Lmatrix}
		\opvec {L} = \mat{C} \opvec{r},
	\end{align}
where we have defined $\mat{C} \coloneqq \begin{pmatrix} \mat{Q} &  \mat{P} \end{pmatrix}  \in \complex^{M \times 2N}$, and the matrices $\mat{Q}$ and $\mat{P}$ (which contain the cooling rates) are comprised of the coefficients in Eq. (\ref{Eq::JumpOps}).  Then, the symmetrized covariance matrix 
obeys the following equations of motion:
	\begin{align}
		\frac{d}{dt} \mat{\Sigma} &= \mat{A} \mat{\Sigma} + \mat{\Sigma} \mat{A}^\tp + \mat{B} , \label{Eq::CovEOM}
	\end{align}
with matrices
	\begin{align}	
		\mat{A} &\coloneqq \mat{\Omega} \bigl[ \mat{G} + \Im \bigl(\mat{C}^\herm \mat{C} \bigr) \bigr], \\
		\mat{B} &\coloneqq \mat{\Omega} \Re \bigl(\mat{C}^\herm \mat{C}\bigr)  \mat{\Omega}^\tp . \label{Eq::EOMCovarianceMatrix}
	\end{align}	
The superscript ${}^\herm$ indicates conjugate transpose of a matrix (to distinguish it from the Hermitian adjoint of an individual operator; see Ref.~\cite{Menicucci2011}).

For the master equation in Eq. (\ref{eq:masterequation}) we have $\mat{G} = \mat{0}$ and
	\begin{align}
		\mat{C}^\herm \mat{C} =   
		\begin{pmatrix}
    			\mat{R}_p & i \mat{T} \\
			-i \mat{T} & \mat{R}_q
		\end{pmatrix},
	\end{align}
where $\mat{T}$, $ \mat{R}_q$, and $ \mat{R}_p$ are symmetric, real matrices. 
The $\mat{A}$ and $\mat{B}$ matrices are block-diagonal:
	\begin{align} \label{Eq::AandBmatrices}
		\mat{A} &=   
		-\begin{pmatrix}
    			 \mat{T} & \mat{0} \\
			 \mat{0} & \mat{T}
		\end{pmatrix} , &
		\mat{B} &=   
		\begin{pmatrix}
    			 \mat{R}_q & \mat{0} \\
			 \mat{0} & \mat{R}_p
		\end{pmatrix}.
	\end{align}
The matrix blocks that comprise $\mat{A}$ and $\mat{B}$ each have a portion corresponding to the CV toric-code cooling map and a diagonal portion corresponding to the local loss,
	\begin{align}
		\mat{T} &= \mat{T}_{\rm TC} + \frac{\gamma_{\rm loss}}{2} \mat \id , \\
		\mat{R}_i &= \mat{R}_{{\rm TC},i} + \frac{\gamma_{\rm loss}}{2} \mat \id, 
	\end{align}
where $i \in \{q,p\}$.
	
Due to the block structure of $\mat{A}$ and $\mat{B}$, the matrix blocks of the covariance matrix that describe quadrature correlations, 
	\begin{align}
		\mat{\Sigma} \coloneqq   
		\begin{pmatrix}
    			 \mat{\Sigma}_{qq} &  \mat{\Sigma}_{qp} \\
			  \mat{\Sigma}_{pq} &  \mat{\Sigma}_{pp}
		\end{pmatrix} ,
	\end{align}
evolve independently. Assuming no $qp$-correlations, $\mat{\Sigma}_{qp} = \mat{\Sigma}_{pq} = \mat 0$, and the diagonal matrix blocks evolve according to
	\begin{subequations} \label{eq::CovBlockEOM}
	\begin{align}
		\frac{d}{dt} \mat{\Sigma}_{qq} =& - \mat{T} \bs{\Sigma}_{qq} - \mat{\Sigma}_{qq} \mat{T} + \mat{R}_{q} , \label{Eq::CovQQEOM} \\
		\frac{d}{dt} \mat{\Sigma}_{pp} =& - \mat{T} \bs{\Sigma}_{pp} - \mat{\Sigma}_{pp} \mat{T} + \mat{R}_{p}  . \label{Eq::CovPPEOM}
	\end{align}
	\end{subequations}
These equations were solved numerically with initial state $\ket \GSvac$ [from Eq.~\eqref{eq:GSvac}] to produce the results presented in Fig.~\ref{fig:mitigate}. There are two facts to note about $\ket \GSvac$. First, it is a Gaussian state since it is the ground state of a quadratic Hamiltonian
\begin{align}
	\op{H}_{\rm SC} = \sum_{i \in \mathcal{V} \cup \mathcal{F}} \op{\eta}^\dagger_i \op{\eta}_i +
	 \op{a}_1^\dagger \op{a}_1
	 +
	  \op{a}_2^\dagger \op{a}_2,
\end{align}
 where $\op{a}_1$ and $\hat{a}_2$ are canonical annihilation operators on the distributed logical modes. Second, $\ket \GSvac$ is an $\mathcal{H}$-graph state~\cite{Menicucci2011} and has no $qp$-correlations. 
	
The matrix-block evolution can be solved analytically %
for the covariance matrix corresponding to the steady-state density matrix $\op{\rho}_{\rm ss}$:
\begin{subequations} \label{eq::CovBlockSteadyState}
	\begin{align}
		\mat{\Sigma}_{qq}(t \rightarrow \infty) = & \frac{1}{2} \mat{T}^{-1} \mat{R}_q, \\
		\mat{\Sigma}_{pp}(t \rightarrow \infty) = & \frac{1}{2} \mat{T}^{-1} \mat{R}_p .  
	\end{align}
	\end{subequations}
In the absence of cooling ($\gamma_{\rm cool} = 0$), the steady state is vacuum. In the opposite regime where there is no loss ($\gamma_{\rm loss}=0$), the steady state is a CV toric-code state, 
$\op{\rho}_{\rm ss} = \ket{\bs{\eta}}\bra{ \bs{\eta} }_{\rm null} \otimes \op{\rho}_\logic$, which depends on the initial state and is in general mixed. When the initial state is the local vacuum, this yields the toric-code vacuum state given by Eq. (\ref{eq:GSvac}), $\hat \rho_{\rm ss} = \ket \GSvac \bra \GSvac$. In the general case both loss and cooling are present, and the steady state is neither pure nor is it a CV toric-code state ($\tr[\op{\eta}_i^\dagger \op{\eta}_i \op{\rho}_{\rm ss} ] \neq 0$ for some or all of the nullifiers). However, for cooling that greatly outweighs loss ($\gamma_{\rm cool}/\gamma_{\rm loss} \gg 1$), the steady state can be close to the CV toric-code vacuum, $\ket \GSvac$, as shown in Fig.~\ref{fig:mitigate}(a). 
\blk

\end{appendix}

\end{document}